\documentclass{article}
\usepackage{graphicx}  
\usepackage{amsmath}   
\usepackage[compress]{cite}
\usepackage{amssymb}   
\usepackage{bm} 
\usepackage{dcolumn}
\usepackage[toc,page]{appendix}
\usepackage{xcolor}
\usepackage{mathrsfs}
\usepackage{amsfonts}
\usepackage{cancel}
\usepackage{varioref}
\usepackage{verbatim}
\usepackage{textcomp}
\usepackage{ tipa }
\usepackage{soul}
\usepackage[normalem]{ulem}
\RequirePackage[colorlinks,citecolor=red,urlcolor=purple,linkcolor=blue]{hyperref}
\allowdisplaybreaks
\labelformat{section}{Section #1} 
\labelformat{subsection}{Section #1} 
\labelformat{subsubsection}{Section #1}
\labelformat{subsubsubsection}{Section #1}
\labelformat{equation}{~(#1)} 
\labelformat{figure}{Fig.~#1} 
\labelformat{subfigure}{Fig.~\thefigure#1} 
\labelformat{table}{Tab.~#1} 
\labelformat{appendix}{Appendix #1}

\def\no{\nonumber}

\def\g{\gamma}
\def\d{\delta}

\def\p{\partial}

\def\na{\nabla}

\def\T{\mathcal{T}}

\def\k{\kappa}

\def\t{\widetilde}

\def\be{\begin{equation}}
\def\ee{\end{equation}}
\def\ba{\begin{align}}
\def\ea{\end{align}}

\def\mg{\sqrt{-g}}

\def\tmg{\sqrt{-\tilde g}}

\def\H{\mathcal{H}}
\def\B{\mathcal{B}}
\def\A{\mathcal{A}}

\def\V{\mathcal{V}}
\def\P{\mathcal{P}}

\def\D{\mathcal{D}}
\labelformat{section}{Section #1} 
\labelformat{subsection}{Section #1} 
\labelformat{subsubsection}{Section #1}
\labelformat{subsubsubsection}{Section #1}
\labelformat{equation}{Eq.~(#1)} 
\labelformat{figure}{Fig.~#1} 
\labelformat{subfigure}{Fig.~\thefigure#1} 
\labelformat{table}{Tab.~#1} 
\labelformat{appendix}{Appendix #1}
\title{\bf Entropic route to Brown-York tensor: A unified framework for null and timelike hypersurfaces}

\author{Krishnakanta
Bhattacharya\footnote{krishnakanta@dubai.bits-pilani.ac.in}$~^{1}$, Bhera Ram\footnote{bhera.ram@iitg.ac.in}$~^{2}$ and Bibhas Ranjan Majhi\footnote{bibhas.majhi@iitg.ac.in}$~^{2}$
\\
$^{1}${\small{Department of General Science}}\\
{\small{BITS Pilani, Dubai Campus, International Academic City, Dubai, United Arab Emirates}}\\
$^{2}${\small{Department of Physics}}\\
{\small{Indian Institute of Technology Guwahati, Guwahati 781039, Assam, India}}}
\begin{document}
\maketitle
\begin{abstract}

Building on Padmanabhan’s entropy functional, originally introduced to derive Einstein’s equations and highlight the emergent nature of gravity, we demonstrate its robustness in a broader context. Using the same entropy density,
we show that the Brown-York tensor is associated with a tangential projection of the entropy-functional polymomentum, thereby providing a common construction applicable to both timelike and null hypersurfaces. This perspective places the two (null and timelike) cases on the same variational footing and clarifies the structural differences of the null Brown-York tensor, including its generally non-symmetric character, without resorting to separate or ad hoc prescriptions in the formulation.  We further extend the analysis to scalar-tensor theories, showing that the entropy-based formulation reproduces the expected equations of motion along with the corresponding BY tensor. Our results provide a coherent variational interpretation of quasi-local gravitational quantities and reveal a common underlying structure linking bulk dynamics and boundary momentum.
\end{abstract}
\section{Introduction}
The formulation of conserved quantities in gravitational theories has long been a subtle and conceptually rich problem. Unlike standard field theories defined on fixed backgrounds, general relativity does not admit a local covariant energy--momentum tensor for the gravitational field itself, primarily due to diffeomorphism invariance and the equivalence principle. Nevertheless, several important notions of gravitational energy and conserved charges have been developed, including ADM \cite{Arnowitt:1960es, Arnowitt:1962hi}, Bondi \cite{Bondi:1962px, Sachs:1962wk}, Komar \cite{Komar:1958wp}, ADT \cite{Abbott:1981ff, Deser:2002rt, Deser:2002jk}, and covariant phase-space constructions \cite{Wald:1993nt,Iyer:1994ys}, most of which are inherently global or quasi-local in nature. Among the various quasi-local approaches \cite{Misner:1964je,Hawking:1968qt, Hayward:1994bu, Hayward:1993ph, Penrose:1982wp}, the Brown--York (BY) tensor provides one of the most widely used and physically meaningful constructions \cite{Brown:1992br,Brown:2000dz}. Defined via the Hamilton-Jacobi variation of the gravitational action with respect to the induced metric on a boundary hypersurface, the BY tensor is often interpreted as the quasi-local stress tensor associated with observers residing on that boundary.
For timelike (or spacelike) hypersurfaces, this construction is well understood. The presence of a non-degenerate induced metric allows one to define the canonical momentum conjugate to the boundary metric, leading directly to the familiar expression for the BY tensor in terms of the extrinsic curvature. This framework also leads to a conservation equation on the boundary, reinforcing the interpretation of the BY tensor as an effective energy-momentum tensor.

However, the situation becomes considerably more subtle for null hypersurfaces. The induced metric on a null surface is degenerate, and hence the standard Hamilton-Jacobi prescription cannot be applied in a straightforward manner. As a result, existing construction of a null analogue of the BY tensor typically rely on auxiliary structures, such as the introduction of an additional (auxiliary) null vector, and involve treating multiple geometric quantities as independent boundary data. Furthermore, the resulting null BY tensor is generically non-symmetric (see \cite{Chandrasekaran:2021hxc,Adami:2023fbm,Bhattacharya:2023ycc, Bhattacharya:2024fbz}), in contrast to its timelike counterpart, raising questions about its precise physical interpretation. These features indicate that the conventional understanding of the BY tensor as a stress tensor may be incomplete or, at best, context-dependent.

In this work, we revisit the origin and interpretation of the Brown-York tensor from a different perspective. We analyze that the BY tensor is more naturally understood as a canonical momentum density associated with boundary data, rather than as a conventional energy-momentum tensor. This viewpoint becomes particularly transparent when one analyzes the structure of boundary terms arising in the on-shell variation of an action, where the conjugate momenta are identified directly from surface contributions. Here we develop a unified variational framework based on an entropy density functional, following the emergent gravity paradigm proposed by Padmanabhan \cite{Padmanabhan:2004kf, Padmanabhan:2007en}. In this approach, spacetime dynamics is obtained by extremizing an entropy functional, and the field equations emerge naturally via the variational principle. A key advantage of this formulation is that it treats normal vector fields as the fundamental variables, rather than the induced metric on the hypersurface, hence it can be applied to null surfaces as well. As we demonstrate, upon projecting the canonical momentum conjugate to these normal vectors on the relevant hypersurface naturally gives rise to the Brown-York tensor.

This entropy-based construction provides a common and systematic route to obtain the BY tensor for both timelike and null hypersurfaces. In particular, it bypasses the difficulties associated with the degeneracy of the induced metric on null surfaces and clarifies the origin of the additional structures required in the standard approach. We further extend our analysis beyond general relativity to scalar–tensor theories of gravity, which provide a natural generalization and encompass models such as $f(R)$ gravity. In this broader setting, we show that the entropy-based variational principle continues to yield the appropriate form of the Brown-York tensor, while also elucidating the origin of its non-conservation: the Brown-York tensor alone is not conserved because of the exchange of momentum between the gravitational and scalar sectors.

The organization of the paper is as follows. In \ref{SECVECFIELD}, we briefly discuss vector field theory in order to motivate the later interpretation of the Brown-York (BY) tensor as a momentum-density tensor, despite its two-index structure. We emphasize that an analogous situation already arises in vector field theory, where the indices carry distinct physical meanings and should not be treated on equal footing. In \ref{BYrevisit}, we revisit the existing formulations of the BY tensor in both general relativity (GR) and scalar-tensor (ST) gravity, highlighting the subtleties associated with null hypersurfaces and modified theories of gravity. In \ref{SECourformalismGR}, we introduce our unified entropy-based formalism within general relativity, in which the BY tensor naturally emerges as the projection of the conjugate momentum associated with the surface normals for both timelike and null boundaries. We then extend the framework to scalar-tensor gravity in \ref{SECSTGRAV}, where the same entropic variational principle simultaneously yields the bulk gravitational field equations together with the generalized BY tensor. Finally, we summarize our results and discuss their implications in \ref{concl}. At the end two appendices are included to provide the explicit steps in deriving few relevant equations.

\textit{Notations and conventions:} We adopt the mostly positive signature for the metric, with the flat metric given by $\eta_{ab} = \mathrm{Diag}(-1, +1, +1, +1)$. Null normals are represented by $l^a$, with the corresponding auxiliary null vectors denoted by $k^a$. For a timelike surface (defined by the normal $s^a$), the induced metric is denoted by $\gamma_{ab}$, and the extrinsic curvature is denoted by $K_{ab}$. A timelike normal is denoted by $n_a$, and a spacelike normal by $s_a$. The null projection tensors and relevant quantities are specified in the appropriate place.
\section{Obtaining insights from vector field theory} \label{SECVECFIELD}
As emphasized in the introduction, our goal is to extract structural insight into the Brown--York (BY) tensor by addressing key issues. We ask if there is any unified approach to obtaining the BY tensor for timelike as well as for the null surface. For timelike hypersurfaces, the BY tensor has a well-defined Hamilton–Jacobi interpretation. In that case, the Brown-York tensor is defined as (on-shell)
\begin{align}
    T_{ab}^{\rm BY}=\frac{-2}{\sqrt{\gamma}}\frac{\p A_{\rm Tot}}{\p \gamma^{ab}}~,\label{bygen}
\end{align}
where $A_{\rm Tot}$ is the total action consisting of the Einstein-Hilbert gravitational action and Gibbons-Hawking-York (GHY) boundary term. However, despite knowing the boundary term for the null surface (and, thereby, the total action), we notice that the null BY tensor cannot be expressed in the form of \ref{bygen}. This is a known challenge, as most works treat timelike BY tensor cleanly, but null BY tensor requires ad-hoc tricks \cite{Chandrasekaran:2021hxc,Adami:2023fbm,Bhattacharya:2023ycc} due to the degeneracy of the induced metric. This stands in contrast with the timelike case and suggests that the underlying object may not be a conventional energy-momentum tensor. Our aim is to show that both of these features admit a natural explanation once the BY tensor is interpreted more fundamentally as a \emph{momentum density}.
To motivate this perspective, we begin with classical mechanics.

In the Lagrangian description of classical mechanics, the Euler-Lagrange equation of motion is obtained by the least action principle. In general, if the action is described as
\begin{equation}
S = \int_1^2 dt ~L(q,\dot{q})~,
\label{B1}
\end{equation}
then its variation under an arbitrary change in $q \to q+\delta q$ is given by
\begin{equation}
\delta S = \int_1^2 dt ~\mathcal{E} ~\delta q + p\delta q\Big|_{1}^{2}~.
\label{B2}
\end{equation}
Here $\mathcal{E}$ is the Euler derivative and $p = \partial L/\partial \dot{q}$ is the canonical momentum. The above variation can provide two important information about the classical system. If one adopts least action principle, in which $\delta q(2) = 0 = \delta q(1)$, and set $\delta S=0$, then the equation governing the dynamics of the system (i.e. $\mathcal{E} = 0$) is obtained. On the other hand if the variation is done on-shell (i.e. along the classical path and so one imposes $\mathcal{E}=0$ in \ref{B2}), then fixing the initial position (i.e. $\delta q(1)=0$), one finds 
\begin{equation}
\delta S = p\delta q\Big{|}_2~.   \label{delspdq}
\end{equation}
Then the canonical momentum is defined from the boundary term in terms of the action as $p = \partial S/\partial q$. This latter definition of canonical momentum, as obtianed in \ref{delspdq}, is important for the Hamilton-Jacobi formalism. In that case, the Hamiltonian is defined as $H = - (\partial S/\partial t)$. For a relativistic particle these are expressed through single relation $p_a = \partial S/\partial x^a$, with $p^a = (H,\vec{p})$ and $x^{a} = (t,\vec{x})$.

This structure carries over to classical field theory, but with an important refinement: momentum is now associated with a hypersurface rather than a single point. A brief discussion on the Maxwell's field will be indicative for our main goal.
The action for the massless vector field $A_i$ is given as:
\begin{align}
    S[A]=-\frac{1}{4}\int_{\V}d^4x~F^{ij}F_{ij}~, ~~~~~~~~~~~~~~~~~\textrm{where}~~~~~~~~~~~~~~~~~ F_{ij}=\p_iA_j-\p_jA_i~. \label{SA}
\end{align}
Upon varying the above action \ref{SA}, one obtains
\begin{align}
    \d S[A]=-\int_{\V}d^4x~F^{ij}(\p_i\d A_j)=\int_{\V}d^4x~(\p_iF^{ij})\d A_j-\int_{\p\V}d\Sigma_i~F^{ij}\d A_j~.
\label{B4}
\end{align}
The last term is the boundary term. If the fields are localized, then the contribution of the boundary term from the timelike surface vanishes, and hence the above reduces to
\begin{equation}
\d S[A]=\int_{\V}d^4x~(\p_iF^{ij})\d A_j-\int_{{\p\V}_t}d\Sigma_i~F^{ij}\d A_j~, 
\end{equation}
where ${{\p\V}_t}$ refers to the $t=$ constant part of the full boundary ${\p\V}$. Now like earlier, if we now fix $\delta A_i$ on ${{\p\V}_t}$ with $\d S[A]=0$ then it yields the field equations
\begin{align}
    \p_iF^{ij}=0~.
 \label{B5}   
\end{align}
On the other hand for the on-shell variation, \ref{B4} reduces to the boundary contribution:
\begin{align}
    \d S[A]=-\int_{\p\V}d\Sigma_i~F^{ij}\d A_j~.
\end{align}
From this boundary term, one may identify the momentum density conjugate to the field variable $A_j$ across a hypersurface normal to the direction $i$ as
\begin{align}
    P^{(i)j}=-F^{ij}=F^{ji}~.
\end{align}
At this stage, an important conceptual point emerges. The momentum density carries two indices, but these indices have different meanings. One index (here it is $i$) specifies the direction normal to the hypersurface (i.e., it tells us which boundary we are considering), while the other (which is $j$ here) labels the corresponding field variable. Therefore these indices are therefore not on equal footing, and there is no fundamental reason for the object to be symmetric.
Hence we call $P^{(i)j}$ as the $i^{th}$ component of the momentum corresponding to $A_j$ field.
For instance, on a spacelike surface with normal along the time direction, the canonical conjugate momentum is
\begin{align}
    P^{(0)j}=\Pi^j=-F^{0j}~,
\end{align}
so that the momentum conjugate to $A_j$	is precisely $-F^{0j}$ which corresponds to the components of the electric field. Furthermore, since this Lagrangian density is not an explicit function of $A_i$, the corresponding canonical momentum $P^{(j)i} = F^{ij}$
 must be conserved; i.e. $\partial_jP^{(j)i} = 0$. This can be easily verified by using the equation of motion \ref{B5}. However, this conservation of $4-$momentum of $A_i$ field is not true when the Lagrangian density is an explicit function of the field itself. 

With this perspective in mind, we will show that the Brown-York tensor has an analogous interpretation. As a result, although it is often referred to as an energy-momentum tensor, it is more accurately understood as a momentum density conjugate to geometric variables defined on a hypersurface. This viewpoint naturally explains both the difficulty in extending the standard definition to null surfaces. Before going into this goal, in the next section, we will briefly mention what the literature says on the Brown-York tensor both for timelike and null-like. This will help us to present the existing issues in a more transparent way to the readers. Additionally, it will act as building blocks for our main discussion in terms of notations as well as various defined variables.  
\section{Brown-York tensor: Subtleties in null boundaries and modified theories} \label{BYrevisit}
\subsection{Within general relativity}
{\it Non-null boundary: --}
In the literature, the Brown–York tensor is obtained from the total gravitational action using the Hamilton–Jacobi variation with fixed boundary metric data. The total action (gravity + Gibbons-Hawking-York + matter) is taken as
\begin{align}
    \mathcal{A}_{\rm tot}=\A_{\rm Grav}+\A_{\rm GHY}+\A_{\rm matter}=\frac{1}{16\pi}\int_{\V}\mg R~d^4x-\frac{\epsilon}{8\pi}\int_{\p\V}\sqrt{h}\mathcal{K}d^3x+\int_{\V}\mg d^4x\mathcal{L}^{\rm (mat)}~,
\end{align}
where $\epsilon=+1$ for timelike $\partial\mathcal{V}$ and $\epsilon=-1$ for spacelike $\partial\mathcal{V}$, and $h_{ab}$ is the induced metric on the boundary with extrinsic curvature $\mathcal{K}_{ab}$ and trace $\mathcal{K}=h^{ab}\mathcal{K}_{ab}$.

Varying the total action yields the bulk Einstein equation term plus boundary terms.
\begin{align}
    \d \mathcal{A}_{\rm tot}=\frac{1}{16\pi}\int_{\V}\mg \Big(G_{ab}-8\pi T^{\rm{(mat)}}_{ab}\Big)\d g^{ab}~d^4x+\epsilon\int_{\p\V}\sqrt{h}\Big({D}_a\mathcal{T}^a+\mathcal{P}_{ab}\d h^{ab}\Big)d^3x~,
\end{align}
where
\begin{align}
    \mathcal{T}^a=\frac{h^a_ir_j\d g^{ij}}{16\pi}~;~~~~~~~~~~~~~~\mathcal{P}_{ab}=\frac{1}{16\pi}(\mathcal{K}h_{ab}-\mathcal{K}_{ab})~;
\end{align}
and ${D}_a$ is the three-derivative on the surface($\partial \mathcal{V}$), which is compatible with $h_{ab}$ (i.e. ${D}_ah_{bc}=0$). The operation of ${D}_a$ is defined for any vector $A^b$ as 
 \begin{align}
     {D}_a A^b=h^i_ah^b_j\nabla_i(h^j_k A^k)~.
 \end{align}
For a general (non-null) surface with the on-shell condition $G_{ab}=8\pi T^{\rm{(mat})}_{ab}$, one can obtain
\begin{align}
    \frac{\d \mathcal{A}_{\rm tot}}{\d h^{ab}}\Bigg|_{\rm{(on-shell)}} = \epsilon \sqrt{h}\mathcal{P}_{ab}~.
\end{align}
So $P_{ab}$ is the canonical momentum conjugate to $h_{ab}$.
Specializing in a timelike boundary with a induced metric $\gamma_{ab}$, the Brown–York stress tensor given by \ref{bygen} reduces to $T_{ab}^{\rm (BY)}=-2 P_{ab}$, where
\begin{align}
    T_{ab}^{\rm (BY)}=-\frac{2}{\sqrt{\gamma}} \frac{\d \mathcal{A}_{\rm tot}}{\d \gamma^{ab}}=\frac{1}{8\pi}\Big(K_{ab}-K\gamma_{ab}\Big)~.
    \label{BYDEF}
\end{align}
Here, $K_{ab}$ is the extrinsic curvature of the timelike boundary and $K=\gamma^{ab}K_{ab}$. Its divergence satisfies the constraint
\begin{align}
    D_a T^{ab}_{\rm (BY)}=-T_{\rm{(mat)}}^{ac}s_a\gamma^b_c~.
\end{align}
where $D_a$ is the covariant derivative intrinsic to the boundary, $s_a$ is the outward-pointing unit normal to the bulk matter stress tensor projection, and $T_{\rm{(mat)}}^{ac}$ the bulk matter stress tensor. In vacuum (absence of matter flux across the boundary) this reduces to the conservation law
\begin{align}
    D_a T^{ab}_{\rm (BY)}=0~.\label{BYCON}
\end{align}
These relations make $T^{ab}_{\rm (BY)}$ play the role of a quasi-local energy-momentum tensor for an observer confined to the boundary. However, the construction relies crucially on the nondegeneracy of the induced metric; a direct analogue of \ref{BYDEF} cannot be written on a null boundary because the induced metric on a null hypersurface is degenerate. 

{\it Null boundary: --} Following Carter's formulation  \cite{Hawking:1979ig}, one can introduce an auxiliary null vector $k^a$, where $l^a$ and $k^a$ satisfy the following relations:
\begin{align}
    l^al_a=0=k^ak_a, ~~~~~~~~~~~~l^ak_a=-1.
\end{align}
Using these, we may define projection operators useful on and transverse to the null surface. One convenient set is
\begin{align}
	\Pi^a_{~b}=\d^a_b+k^al_b~,
	\no 
	\\
	q^a_b=\d^a_b+l^ak_b+k^al_b~.
\end{align}
Here $\Pi^a_{~b}$ is only a projection tensor (and not an induced metric), projecting the quantities on the null surface $\H$, with the following properties:
\begin{align}
	\Pi^a_{~b}l^b=l^a~,~~~\Pi^a_{~b}l_a=0~,~~~\Pi^a_{~b}k^b=0~,~~~\Pi^a_{~b}k_a=k_b~,
	\no 
	\\
	\Pi^a_{~b}\Pi^b_{~c}=\Pi^a_{~c}~.~~~~~~~~~~~~~~~~~~~~~~~~~~ \label{PIABCON}
\end{align}
On the contrary, $q^a_b$ is the projection tensor onto the two-surface $\B^{\rm (null)}$ upon which both $l^a$ and $k^a$ are the normals (i.e. cross-section of $\H$). The properties of $q^a_b$ are as follows:
\begin{align}
	\mathbf{q}\cdot \mathbf{l}=\mathbf{q}\cdot \mathbf{k}=0~,
	\no 
	\\
	q^a_bq^b_c=q^a_c~.
\end{align}
Unlike $\Pi^a_{~b}$, $q^a_b$ is the induced metric on $\B^{\rm (null)}$. 

A commonly used form of the total action with a null boundary $\mathcal{H}$ replaces the GHY term by an appropriate null boundary term (choices vary in the literature \cite{Parattu:2015gga,Parattu:2016trq,Chakraborty:2017zep,Chakraborty:2018dvi,Lehner:2016vdi,Hopfmuller:2016scf,Oliveri:2019gvm,Aghapour:2018icu,Chandrasekaran:2020wwn}). One convenient choice is \cite{Parattu:2015gga}
\begin{eqnarray}
    &&\A_{\rm tot}^{\rm (null)}=\A_{\rm Grav}+\A_{\rm boundary}^{\rm (null)}+\A_{\rm matter}
    \no\\
    &&~~~~~~~~~~=\frac{1}{16\pi}\int_{\V}\mg R~d^4x+\frac{1}{8\pi}\int_{\H}\mg\Big(\theta+\k\Big)d^3x+\int_{\V}\mg d^4x\mathcal{L}^{\rm (mat)}~, \label{actnulltot}
\end{eqnarray}
where $\theta$, also known as the expansion scalar, is the trace of the second fundamental form $\theta_{ab}$ and $\kappa$ is the nonaffinity (surface gravity). $\theta_{ab}$, $\theta$, and $\kappa$ are defined as follows: 
\begin{align}
    \theta_{ab} =  q_{a}^{~i} q_{b}^{~j} \nabla_i l_j~;~~~~~~~~~~~~~~~~~~~~~~~~\theta=q^{ij}\nabla_i l_j~;~~~~~~~~~~~~~~~~~~~~~~~~l^b \nabla_b\ l^a = \k l^a ~.
\end{align}
Upon varying the total action given by \ref{actnulltot}, one obtains the following \cite{Parattu:2015gga}
\begin{eqnarray}
    \d\A_{\rm tot}^{\rm (null)}&=&\frac{1}{16\pi}\int\mg \Big(G_{ab}-8 \pi T^{\rm{(mat)}}_{ab}\Big)\d g^{ab}\ d^4x
    \nonumber
    \\
    &+&\int_{\H}\Big(\frac{1}{16\pi}\p_a(\mg\Pi^a_{~b}\d l^b_{\bot})+\mg(\P^{\rm (\mathbf{q})}_{ab}\d q^{ab}+\P^{(\mathbf{l})}_a\d l^a)\Big)d^3x~, \label{nullvariation}
\end{eqnarray}
where $l^a_{\bot}=\d l^a+g^{ab}\d l_b$. In addition, $\P^{\rm (\mathbf{q})}_{ab}$ and $\P^{(\mathbf{l})}_a$ are the conjugate momenta corresponding to $q^{ab}$ and $l^a$ respectively, which are given as
\begin{align}
\P^{\rm (\mathbf{q})}_{ab}=\frac{1}{16\pi}\Big(\theta_{ab}-(\theta+\kappa)q_{ab}\Big)~,~~~~~~~~~~~\P^{(\mathbf{l})}_a=\frac{1}{8\pi}\Big((\theta+\kappa)k_a+\omega_a\Big)~,\label{nullmoms}
\end{align}
with $\omega_a$ is the twist 1-form (connection on the normal bundle) defined in the usual way:
\begin{align}
	\omega_a = -\Pi_a^{~b} k^c \nabla_b l_c = -k^b \nabla_a l_b - l_a (k^b k^c \nabla_b l_c) = l^b \nabla_b k_a ~.
	\label{rotnoneform}
\end{align} 
These expressions, \ref{nullvariation} and \ref{nullmoms}, encode how variations of the transverse 2-metric $q^{ab}$ and the null generator $l^a$ appear as boundary data on a null hypersurface. As a result, one must fix both $\delta q^{ab}$ and $\delta l^a$ on the boundary. Consequently, the null Brown–York (BY)–like tensor cannot be obtained directly in the form of \ref{BYDEF}. Instead, since both $q^{ab}$ and $l^a$ serve as dynamical variables, the null BY–like tensor arises indirectly from their conjugate quantities, $\mathcal{P}^{\rm (\mathbf{q})}_{ab}$ and $\mathcal{P}^{(\mathbf{l})}_a$, and is proposed as \cite{Chandrasekaran:2021hxc,Adami:2023fbm,Bhattacharya:2024fbz}
\begin{align}
T^{a\rm{(BY)}}_{~b\rm (null)}=2q^{ai}\P^{\rm (\mathbf{q})}_{ib}+l^a\P^{(\mathbf{l})}_b=\frac{1}{8\pi}\Big(W^a_{~~b}-\Pi^a_{~b} W\Big)~, \label{BYnull}
\end{align}
where $W^a_{~~b}=\theta^a_b+l^a\omega_b$ and $W=\theta+\kappa$~. 

The structure of ~\ref{BYnull} closely resembles the Brown–York tensor defined on a timelike surface, with the correspondence $K^a_{~b}\to W^a_{~~b}$ and $h^a_{~b}\to \Pi^a_{~b}$. Moreover, one can verify that the null BY–like tensor defined in \ref{BYnull} satisfies the following conservation-like relation:
\begin{align}
\D_aT^{a\rm{(BY)}}_{~b\rm (null)}=T^{\rm{(mat)}}_{ab}\Pi^a_{~c}l^b~, \label{connullBY}
\end{align}
for a covariant derivative $\D_a$ defined through $\Pi^a_{~b}$ {\footnote{Such a covariant derivative has been introduced in \cite{Chandrasekaran:2021hxc}. However, since $\Pi^a_{~b}$ is not a projector on the null hypersurface, the definition may be considered as ad-hoc.}}.
Again, in the absence of a matter field, it boils down to the conservation of the null BY tensor $\D_aT^{a\rm{(BY)}}_{~c\rm (null)}=0$.
Thus, although for a null surface the Brown–York tensor cannot be defined through a direct analogy with the standard energy–momentum tensor, it can nevertheless be obtained indirectly. Importantly, it still satisfies a conservation relation analogous to that of the energy–momentum tensor. 

It must be emphasized that the formulation of the Brown--York tensor on null hypersurfaces remains, within the current literature, somewhat ad hoc. Unlike the timelike case, where the BY tensor follows naturally from a well-defined Hamilton--Jacobi prescription through variation of the action with respect to the induced metric, the null construction lacks such a systematic variational foundation due to the degeneracy of the null induced metric. Consequently, the null BY tensor is typically introduced as a suitable combination of conjugate quantities arranged so as to satisfy a conservation-like relation on the null boundary.
Furthermore, the situation is complicated by the fact that multiple inequivalent expressions for the null BY tensor currently exist in the literature. This ambiguity itself reflects the absence of a unique and universally accepted formulation for null boundaries. For example, alternative expressions for the null BY tensor have been proposed in Ref.~\cite{Jafari:2019bpw}, leading to differing interpretations of the associated quasi-local quantities. These issues strongly suggest that the null case requires a more fundamental and unified variational framework, which is one of the primary motivations behind the present work.

\subsection{Scalar-tensor gravity}
Now, the situation becomes more interesting when we look beyond general relativity. Let us take the example of scalar-tensor (ST) gravity, which is an extension of GR. Moreover, the higher-order $f(R)$ gravity can also be studied as a special case of ST gravity. The total action for ST gravity in the Jordan frame, including the appropriate boundary term, is given by \cite{Bhattacharya:2023ycc}
\begin{eqnarray}
&&\mathcal{A}_{\rm J}=\A^{\rm (grav)}[\boldsymbol{g},\phi]+\A^{\rm(mat)}[\boldsymbol{g},\psi]+\A^{\rm (boundary)}
\no\\
&& =\int_{\mathcal{V}} d^4x\sqrt{-g}\Big[L^{\rm (grav)}(\boldsymbol{g},\phi)+L^{\rm(mat)}(\boldsymbol{g},\psi)\Big]+\int_{\p\V}d^3x\sqrt{h}L^{\rm boundary}(\boldsymbol{g},\phi)~,
\no 
\\
&& =\int_{\mathcal{V}} d^4x\sqrt{-g}\Bigg[\frac{1}{16\pi}\Big\{\phi R-\frac{\omega(\phi)}{\phi}g^{ab}\nabla_a\phi \nabla_b\phi-V(\phi)\Big\}+L^{\rm(mat)}(\boldsymbol{g},\psi)\Big]-\frac{\epsilon}{8\pi}\int_{\p\V}\sqrt{h}\phi\mathcal{K}d^3x~,
\label{SJ}
\end{eqnarray}
where $\phi$ is the Brans-Dicke scalar, $\omega(\phi)$ is the  ``Brans-Dicke coupling", and $V(\phi)$ is the scalar field potential.
On varying the above action \ref{SJ}, one obtains \cite{Bhattacharya:2023ycc}
\begin{align}
    \delta \A_{\rm J}&=\int_{\mathcal{V}} d^4x\sqrt{-g}\left(E_{ab}\delta g^{ab}+E_{\phi}\delta\phi\right)+\epsilon\int_{\p\V}\sqrt{h}\Big[D_a\mathcal{T}^a(\boldsymbol{h},\phi)+\P_{ab}(\boldsymbol{h},\phi)\d h^{ab} + \mathcal{P}^{\rm{(\phi)}}\delta \phi\Big]d^3x~,
\end{align}
where the equations of motion of the metric tensor and the scalar field are given as
\begin{align}
E_{ab}&\equiv \frac{1}{16\pi}\Big[\phi G_{ab}+\frac{\omega}{2\phi}\nabla_i\phi\nabla^i\phi g_{ab}-\frac{\omega}{\phi}\nabla_a\phi\nabla_b\phi
+\frac{V}{2}g_{ab}-\nabla_a\nabla_b\phi+\nabla_i\nabla^i\phi g_{ab}-8\pi T^{\rm{(mat)}}_{ab}\Big]=0~,
\label{EoMST}
\\
E_{\phi}&\equiv \frac{1}{16\pi}\Big[ \square\phi-\frac{1}{2\omega+3}\Big(8\pi T^{\rm{(mat)}}-\frac{d\omega}{d\phi}\nabla_i\phi\nabla^i\phi+\phi \frac{dV}{d\phi}-2V\Big)\Big]=0~.\label{EPHIJOR}
\end{align}
In addition, the expressions for $\mathcal{T}^a(\boldsymbol{h},\phi)$, $\P_{ab}$ and $\mathcal{P}^{\rm{(\phi)}}$ are given as \cite{Bhattacharya:2023ycc}
\begin{align}
    \mathcal{T}^a(\boldsymbol{h},\phi)=\frac{\phi h^a_ir_j\d g^{ij}}{16\pi}~;~\P_{ab}(\boldsymbol{h},\phi)=\frac{1}{16\pi}\Big[\phi(\mathcal{K}h_{ab}-\mathcal{K}_{ab})-h_{ab}r^i\na_i\phi\Big]~;~ \mathcal{P}^{\rm{(\phi)}} = -\frac{1}{8\pi}\Big(\mathcal{K} + \frac{\omega}{\phi}r^i\na_i \phi\Big)~.
\end{align}

From this, the gravitational momentum conjugate to the induced metric on a timelike surface is
\begin{align}
    P_{ab}(\boldsymbol{h},\phi)=\frac{1}{16\pi}\Big[\phi(K\gamma_{ab}-K_{ab})-\gamma_{ab}s^i\na_i\phi\Big]~.
\end{align}
Defining Brown-York tensor(for timelike surface) as $T_{ab}^{\rm (BY)}(\boldsymbol{\gamma},\phi)=-2 P_{ab}(\boldsymbol{\gamma},\phi)$, one obtains
\begin{align}
    T_{ab}^{\rm (BY)}(\boldsymbol{\gamma},\phi)=-\frac{2}{\sqrt{\gamma}}\frac{\d \mathcal{A}_{\rm J}}{\d \gamma^{ab}}=\frac{1}{8\pi}\Big[\phi\Big(K_{ab}-K\gamma_{ab}\Big)+\gamma_{ab}s^i\na_i\phi\Big]~. \label{STBY1}
\end{align}

Furthermore, the null BY tensor has also been obtained recently in \cite{Bhattacharya:2023ycc}.
It is important to note that, in contrast to GR, the Brown-York tensor in scalar-tensor gravity does not satisfy the usual conservation relation. Instead, as shown in \ref{App2} (also see \ref{appnull} for the null case), we obtain 
\begin{align}
    D_a T^{ab}_{\rm (BY)}(\boldsymbol{\gamma},\phi)=-T_{\rm{(mat)}}^{ac}s_a\gamma^b_c + \gamma^{ab}\Pi^{(\phi)}D_a \phi~.
    \label{Conservation1}
\end{align}
where, $\Pi^{(\phi)} =  -\frac{1}{8\pi}\big(K+ \frac{\omega}{\phi}s^c \na_c \phi\big)$ is the conjugate quantity $\mathcal{P}^{\rm{(\phi)}}$ on a timelike hypersurface \cite{Bhattacharya:2023ycc}. 
This implies that the usual conservation law is modified by the flux of the scalar field and its associated momentum across the boundary. Hence its non-conservation should not be interpreted as a violation of the underlying momentum balance; rather, it reflects the exchange of momentum between the gravitational boundary degrees of freedom and the scalar sector.

Interpreting the BY tensor as a conventional ``energy-momentum tensor'' therefore requires some care. Rather, it is more naturally understood as a quasi-local boundary momentum or response tensor, whose energy--momentum interpretation emerges only after an appropriate projection and choice of boundary observers. This becomes particularly transparent in the entropic variational framework developed in this work. 
As a side remark, the interpretation of the BY tensor as a momentum density rather than a symmetric energy-momentum tensor also clarifies another important feature: its lack of symmetry on null hypersurfaces. Since the two indices of the BY tensor encode different geometric roles, one associated with the hypersurface normal and the other with the choice of field variable, they are not required to be interchangeable. This asymmetry is therefore not an anomaly but a direct reflection of its underlying canonical structure \footnote{This is similar in nature with the idea of usual Noether prescription to find the energy-momentum tensor. Usually Noether prescription, corresponding to translational invariance, provides symmetric energy-momentum tensor for free scalar field while it fails to do same for the electromagnetic case. However, using a certain trick, like Belinfante tensor for electromagnetic case, one can symmetrize those obtained from Noether prescription. On the other hand the definition of energy-momentum tensor through the variation of the action with respect to the metric tensor always confirms the symmetric property.}. Consistently, the standard Hamilton-Jacobi prescription, which relies on variations with respect to a non-degenerate induced metric, fails to capture the null case, where such a metric does not exist. With this discussion let us now go into the main analysis, which will illuminate various questions addressed till now.


\section{Entropy density to Brown-York tensor} \label{SECourformalismGR}
In the previous section, we observed that existing methods for obtaining the Brown-York (BY) tensor on timelike or null surfaces lack uniformity. In particular, the null BY tensor cannot be constructed in the standard way because the null induced metric is degenerate, and it cannot be derived from the usual Hamilton–Jacobi principle. The ad hoc formulation for null surfaces \cite{Chandrasekaran:2021hxc,Adami:2023fbm,Bhattacharya:2023ycc} creates an apparent asymmetry in how the BY tensor is defined across different types of hypersurfaces.

In this section, we present an alternative and unified route to obtain the BY tensor in both general relativity and scalar–tensor gravity, a natural extension of GR. Our approach is based on entropy density, following the {\it emergent gravity} perspective originally proposed by Padmanabhan \cite{Padmanabhan:2004kf, Padmanabhan:2007en}, who demonstrated that Einstein’s equations can be obtained by extremizing an appropriate entropy functional {\footnote{An extensive discussion on emergent nature of gravity and the construction of the entropy functional in this paradigm are beyond the scope of this paper. For detailed ideas on this direction, refer to chapter 16 of \cite{Padmanabhan:2010zzb}.}}. In this picture, spacetime dynamics is not introduced axiomatically but emerges from a thermodynamic variational principle. Building on this idea, we show that the same entropy-based formulation not only reproduces the BY tensor in general relativity {\footnote{In fact, the possibility of obtaining the Brown-York tensor as a conjugate quantity emerging from the entropy density was somewhat implicit in the works of T. Padmanabhan and collaborators \cite{Padmanabhan:2010rp, Kolekar:2011gw}. However, this idea had not previously been investigated in a systematic and rigorous manner, particularly in the context of a unified treatment of both timelike and null hypersurfaces, as developed in the present work. Furthermore, the null BY tensor was not defined back then.}} in a unifying manner for both timelike and null surfaces but also provides a natural route to extend its definition beyond GR. In particular, when applied to scalar–tensor theories, the formalism yields both the gravitational field equations and the corresponding BY tensor in a unified manner. This simultaneous emergence highlights the robustness of the entropy-based approach and clarifies the structural origin of boundary momentum in a broader class of gravitational theories.

The key feature of this approach is that the BY tensor appears as the tangential projection of the canonical momentum associated with the normal vectors to a hypersurface, rather than as a quantity derived indirectly from the action principle. This interpretation provides a consistent treatment for both timelike and null surfaces, thereby overcoming the difficulties associated with the degeneracy of the induced metric on null hypersurfaces.
We first establish the formalism in general relativity and later extend it to scalar–tensor theories.

\subsection{General Relativity}
We begin with the case of a timelike surface and show that the idea provides a consistent result. Inspired by this, the analysis is being then extended to null surfaces.
\subsubsection{Timelike Surface}
The induced metric on a timelike surface is defined as
\begin{align}
    \g_{ab}=g_{ab}-s_as_b~,
\end{align}
where $s_a$ is the unit spacelike normal (i.e. $s^as_a=1$). This projection removes the normal direction and defines the intrinsic geometry of the hypersurface.

The entropy functional for a timelike surface, denoted $S[s]$, is postulated as
\begin{align}
    S[s]=\frac{1}{16\pi}\int_{\V}d^4x\mg s[s] ; ~~~~~~~~\textrm{where}~~~~~~~~~~~s[s]=-\Big(4P^{cd}_{ab}\na_cs^a\na_ds^b-\mathcal{T}_{ab} s^as^b\Big); \,\,\,\ \nabla_i P^{cd}_{ab}=0~.\label{EDTGR}
\end{align}
with $\nabla_a \mathcal{T}^{ab}=0$.
Here, $s[s]$ denotes the entropy density associated with the normal vector field $s^a$. The structure of this expression is motivated by Padmanabhan’s emergent gravity paradigm \cite{Padmanabhan:2004kf, Padmanabhan:2007en}, where it was shown that the entropy can be obtained on-shell from the surface part of the action when the bulk part vanishes due to the equation of motion. In the above action, the first term encodes purely geometric information (depending on derivatives of the normal), while the second term $\mathcal{T}_{ab}$, just a second-order tensor to begin with, will later be shown to be related to the matter energy-momentum tensor.
Furthermore, within GR, $P^{cd}_{ab}$ is defined as 
 \begin{align}
     P^{cd}_{ab}=\frac{1}{2}\Big(\d^c_a\d^d_b-\d^c_b\d^d_a\Big)~.
 \end{align}

 Following Padmanabhan’s method, we vary the entropy functional with respect to the normal vector field $s^a$. Keeping the condition $s^a s_a=1$ in mind, the variations are done subject to the constraint $\delta(s^a s_a)=0$. To enforce this, we introduce a Lagrange multiplier $\Lambda$, yielding the variation
\begin{align}
  &&  \d S[s]=-\frac{1}{8\pi}\int_{\V}d^4x\mg\Big[4P_{ab}^{cd}(\na_cs^a)(\na_d\d s^b)-\T_{ab}s^a\d s^b-\Lambda g_{ab}s^a\d s^b\Big]
  \no 
  \\
  && \equiv -\frac{1}{8\pi}\int_{\V}d^4x\mg\Big[4P_{ab}^{cd}(\na_cs^a)(\na_d\d s^b)-\bar\T_{ab}s^a\d s^b\Big]~,
\end{align}
where $\bar\T_{ab}=\T_{ab}+\Lambda g_{ab}$. Then, integrating by parts and using $\na_iP^{cd}_{ab}=0$, one obtains
\begin{align}
    \d S[s]=\frac{1}{8\pi}\int_{\V}d^4x\mg\Big[4P_{ab}^{cd}(\na_d\na_cs^a)+\bar\T_{ab}s^a\Big]\d s^b+\int_{\p\V}d\Sigma_a t^a_{~b}\d s^b
\end{align}
where the surface term is expressed in terms of the quantity
\begin{align}
    t^i_{~j}=-\frac{1}{2\pi}P^{im}_{jn}\na_ms^n=\frac{1}{4\pi}\Big[\na_js^i-\delta^i_j\na_as^a\Big]~.
    \label{tijBY}
\end{align}

Following the earlier idea, this can be interpreted as the $i^{th}$ component of the canonical momentum conjugate to $s^j$ field.
It can also be noted that $t^i_{~j}$ can also be defined as
\begin{align}
    t^i_{~j}=\frac{1}{16\pi}\frac{\p s[s^a]}{\p(\na_is^j)}~,
\end{align}
which plays the role of the canonical momentum conjugate to $s^a$.
Next, introducing the extrinsic curvature of the timelike surface, $K_{ij}=-\g^a_i\na_as_j=-\na_is_j+s_is^a\na_as_j$, we obtain
\begin{align}
    t^i_{~j}=\frac{1}{4\pi}\Big[-K^i_j+s_js^a\na_as^i+\d^i_j K\Big]~.
\end{align}
The term proportional to $s_j$ does not contribute upon projection onto the hypersurface. Therefore, projecting the canonical momentum onto the hypersurface and fixing the normalization by comparison with the standard Brown-York definition, we obtain
\begin{align}
    T^a_{b~\rm (BY)}=-\frac{1}{2}\g^a_i\g^j_bt^i_{~j}=\frac{1}{8\pi}[K^a_b-\g^a_b K]~,
    \label{TabBY}
\end{align}
which exactly corresponds to the Brown-York tensor. Thus, within the entropy-density-based formulation, the Brown-York tensor is associated with an appropriate tangential projection of the canonical momentum conjugate to the hypersurface normal. The tensorial structure of the Brown--York quantity therefore follows directly from the canonical momentum, while its overall normalization is fixed by requiring agreement with the established Brown-York convention.
Our approach provides a clear and physically motivated interpretation that holds uniformly across timelike and (as we will see later) null surfaces.

Furthermore, Einstein's equation can be obtained from the bulk term with the following arguments. When $\delta s^b$ is kept fixed on the boundary, one obtains
\begin{align}
4 P_{ab}^{cd} \nabla_d \nabla_c s^a + \bar{\mathcal{T}}_{ab} s^a = 0~.
\end{align}
Expanding the above expression, one can obtain
\begin{align}
    R_{ab}=\frac{1}{2}\bar{\mathcal{T}}_{ab}~,
\end{align}
One can then use the Bianchi identity to identify $\bar{\mathcal{T}}_{ab}$ with the matter energy-momentum tensor up to trace terms, giving Einstein’s equation
\begin{align}
    G_{ab}:=R_{ab}-\frac{1}{2}g_{ab}R=\frac{1}{2}\bar{\mathcal{T}}_{ab}-\frac{1}{4}\bar{\mathcal{T}}g_{ab} \label{einnull}
\end{align}
Using Bianchi identity, one can obtain $\na_aT^{ab}=0$, where
\begin{align}
    T_{ab}=\frac{1}{16\pi}\Big(\bar{\mathcal{T}}_{ab}-\frac{1}{2}\bar{\mathcal{T}}g_{ab}\Big) - \frac{1}{8\pi}\Lambda g_{ab}~.
\end{align}
Substituting the above identification in \ref{einnull}, one can obtain
\begin{align}
    G_{ab} - \Lambda g_{ab}:=R_{ab}-\frac{1}{2}g_{ab}R - \Lambda g_{ab}=8\pi T_{ab}~.
\end{align}
In summary, the entropy functional approach yields the Brown–York tensor as the projection of the canonical momentum conjugate to the hypersurface normal and reproduces Einstein’s equations from the bulk variation.
Furthermore, in order to satisfy the Bianchi identity, we would end up with $\Lambda$ being at most a constant if not zero, which is consistent with Einstein's theory.

A few comments on $t^i_{~j}$ are in order. One can check from \ref{tijBY}, one readily obtains
\begin{align}
    \nabla_i t^i_{~j} = \frac{1}{4\pi}R_{ij}s^i \label{CONtij}
~.\end{align}
 Therefore, in vacuum with $\Lambda=0$ solutions, $t^i_{~j}$ is covariantly conserved on-shell. In the presence of a cosmological constant, with $T_{ab}=0$, the Einstein equations instead give 
\begin{equation}
\nabla_i t^i_{~j} = -\frac{1}{4\pi}\Lambda s_j~.
\label{PRDV21}
\end{equation}
Thus, although the full tensor ($t^i_{~j}$) is not covariantly conserved in this case, its divergence is entirely along the hypersurface normal.

It can be noted that different projections of the relation $\nabla_i t^i_{~j}$ (as given in \ref{CONtij}) admit distinct physical interpretations. In particular, the projections $s^j\nabla_i t^i_{~j}$ and $\gamma^j_b\nabla_i t^i_{~j}$ can respectively be identified with the timelike Raychaudhuri equation \cite{Gourgoulhon:2005ng, Poisson:2009pwt} and the energy and momentum constraint equations of the ADM decomposition \cite{Arnowitt:1960es, Arnowitt:1962hi}. A detailed textbook discussion of these relations may be found in \cite{Padmanabhan:2010zzb}.

Let ${D}_a$ be the metric-compatible covariant derivative (i.e. ${D}_a\gamma^a_b = 0$). Then one can show, using \ref{TabBY}, that ${D}_a T^a_{b~\rm{(BY)}} = -\frac{1}{8\pi}\gamma^a_b s^c R_{ac}$. Hence, we have ${D}_a T^a_{b~\rm{(BY)}} =-\frac{1}{2}\gamma^j_b \nabla_i t^i_j $. This relation makes clear that conservation of the Brown-York tensor on the hypersurface is a distinct condition from the covariant conservation of the full canonical momentum $t^i_j$. In particular, when \ref{PRDV21} holds, i.e. the divergence of $t^i_j$ is purely normal to the hypersurface, its tangential projection vanishes yielding the conservation of BY tensor. The Brown-York tensor is therefore conserved on the hypersurface even though the full $t^i_j$ is not covariantly conserved in all spacetime directions. This distinction is important: the conservation of the Brown-York tensor probes only the tangential component of the canonical momentum balance, whereas $\nabla_i t^i_j=0$ constitutes the stronger statement of full covariant conservation. The latter requires both the vacuum condition and a vanishing cosmological constant, while conservation of the Brown-York tensor requires only the vacuum Einstein equations.  

\subsubsection{Null Surface}
We now extend the entropy-based formalism to null hypersurfaces, which requires special care due to the degeneracy of the induced metric. Let $\H$ be a null hypersurface with null normal $l^ al_a=0$. The entropy functional associated with the null surface is given by \cite{Padmanabhan:2004kf,Padmanabhan:2007en,Kolekar:2011gw} 
\begin{align}
    S[l]=\frac{1}{16\pi}\int_{\V}d^4x\mg s[l] ; ~~~~~~~~~~~~~~~~~~~~\textrm{where}~~~~~~~~~~~~~~~~~~~~s[l]=-\Big(4P^{cd}_{ab}\na_cl^a\na_dl^b-\T'_{ab}l^al^b\Big)~.
    \label{SL}
\end{align}
Here, $s[l]$ denotes the entropy density associated with the null normal $l^a$ analogous to the timelike case.
This form of the entropic functional is well motivated by the physical principles, specifically macroscopic symmetries and translational invariance from the theory of elasticity of solids. Just like in solids, the key quantity is the displacement vector field $\xi^a(x)$, which describes the elastic displacement of the solid under $x^a\to x^a + \xi^a(x)$, where all thermodynamic potentials, including entropy, can be written as an integral over a quadratic functional of the displacement vector field, capturing the long wavelength limit of the microscopic theory. Treating the spacetime as a deformable solid with elastic properties, the entropic action is an integral over the spacetime volume to maintain the extensivity of the entropy. Further, just like in the case of elastic solids, we expect the entropy density to be translationally invariant and hence depend only upon the derivatives of the deformation field $\xi^a(x)$. In the context of spacetime solid, the deformation corresponds to the normal vector field of a timelike/spacelike or null hypersurfaces. The lowest order translationally invariant scalar constructed out of the derivatives of $\xi^a(x)$ should have a form $P_{ab}^{~~cd} \nabla_c \xi^a \nabla_d \xi^b$, where the fourth rank tensor is purely built out of the metric and its derivative, maintaining its tensorial nature (therefore out of metric, curvature and its higher order covariant derivatives). But in the presence of matter and other fields, the translational invariance is lost, which is incorporated by adding terms which breaks the translational invariance constructed out of $\xi^a(x)$ explicitly given by $\T'_{ab}\xi^a\xi^b$ to the lowest order. Further, the explicit tensor components $P^{abcd}$ and $\mathcal{T}'_{ab}$ are not free parameters finely tuned by hand; instead, they are uniquely fixed by the single, elegant conceptual demand that any arbitrary coordinate deformation must be thermodynamically permissible as long as the background spacetime obeys the laws of gravitation.
The tensor $\T'_{ab}$, only a two-ranked tensor to begin with (satisfying $\nabla_a\mathcal{T}'^{ab}=0$), will later be identified with the matter energy-momentum tensor on the null surface. In the above we choose $\xi^a=l^a$.

To obtain the equations of motion and the canonical momentum, we vary $l^a\rightarrow l^a+\d l^a$ subject to the null constraint $\d (l^al_a)=0$. Introducing a Lagrange multiplier $\Lambda'$ for the constraint, the variation reads
\begin{align}
  &&  \d S[l]=-\frac{1}{8\pi}\int_{\V}d^4x\mg\Big[4P_{ab}^{cd}(\na_cl^a)(\na_d\d l^b)-\T'_{ab}l^a\d l^b-\Lambda' g_{ab}l^a\d l^b\Big]
  \no 
  \\
  && \equiv -\frac{1}{8\pi}\int_{\V}d^4x\mg\Big[4P_{ab}^{cd}(\na_cl^a)(\na_d\d l^b)-\bar\T'_{ab}l^a\d l^b\Big]~,
\end{align}
where $\bar\T'_{ab}=\T'_{ab}+\Lambda' g_{ab}$. Integrating by parts, the variation can be separated into a bulk and a boundary term:
\begin{align}
    \d S[l]=\frac{1}{8\pi}\int_{\V}d^4x\mg\Big[4P_{ab}^{cd}(\na_d\na_cl^a)+\bar\T'_{ab}l^a\Big]\d l^b+\int_{\H}d^2x\sqrt{q}d\lambda l_a~ t^a_{~b\rm (null)}\d l^b~,
\end{align}
where $\sqrt{q}d^2x$ is the area element of the 2-dimensional cross-section of the null hypersurface, and $\lambda$ parametrizes the null generators. The $i^{th}$ component of the canonical momentum conjugate to $l^j$ can be defined as
\begin{align}
    t^i_{~j\rm{(null)}}  =\frac{\p s[l]}{\p(\na_i l^j)}=-\frac{1}{2\pi}P^{im}_{jn}\na_ml^n=\frac{1}{4\pi}\Big[\na_jl^i-\delta^i_j\na_al^a\Big]~.
\end{align}

Decomposing the derivative of the null normal using the standard null congruence decomposition (see Eq. (5.20) of \cite{Gourgoulhon:2005ng}),
\begin{align}
    \na_jl^i=\theta^i_j+l^i\omega_j-l_jk^m\na_ml^i
    \label{nullcon}
\end{align}
where $\theta^i_j$ is the expansion–shear tensor, $\omega_j$ is the twist/rotation 1-form one can obtain
\begin{align}
    t^i_{j\rm{(null)}}=\frac{1}{4\pi}\Big[\theta^i_j +l^i\omega_j-l_jk^m\na_ml^i-\d^i_j(\theta+\kappa) \Big]~.
\end{align}
Projecting $t^i_{j\rm{(null)}}$ using the null projection tensor $\Pi^a_{~b}$, and fixing the normalisation by maintaing agreement with literature, one obtains the null BY tensor \footnote{Note that for timelike surface, the normalisation factor is taken as $-\frac{1}{2}$, whereas for the null surface it is taken as $\frac{1}{2}$. This difference originates from the sign convention adopted in the definitions of the extrinsic curvature $K_{ij}$ for timelike hypersurfaces and the null extrinsic curvature $\theta_{ij}$ for null hypersurfaces. Specifically, the timelike extrinsic curvature is defined as $K_{ij} = -\gamma^{m}_{~i}\gamma^{n}_{~j}\,\nabla_m n_n$. In contrast, the null extrinsic curvature is defined without the additional minus sign i.e. $\theta_{ij} = q^{m}_{~i}q^{n}_{~j}\nabla_m l_n$. Therefore, the relative sign between the timelike and null Brown-York tensors follows directly from these standard geometric definitions, which explains the different normalization factors appearing in our analysis.}
\begin{align}
    T^a_{~~b\rm{(null)}}=\frac{1}{2}\Pi^a_{~i}\Pi^j_{~b}t^i_{j\rm{(null)}}=\frac{1}{8\pi}\Big[\theta^a_b+l^a\omega_b-\Pi^a_{~b}(\theta+\kappa)\Big]~.
\end{align}
This expression coincides with the null BY tensor found in the literature \cite{Chandrasekaran:2021hxc,Adami:2023fbm}, providing a physically consistent interpretation as the projected part of the canonical momentum conjugate to $l^a$. It, thereby, validates the present entropy-based canonical momentum interpretation. Again, Einstein's equation of motion can be obtained from the bulk part, which can be referred from \cite{Padmanabhan:2004kf, Padmanabhan:2007en}. Thus, in both cases (timelike surface as well as null surface) the bulk term enforces Einstein’s dynamics, while the surface term corresponds to the Brown–York tensor. This dual role reflects the emergent gravity paradigm: the spacetime field equations emerge as consistency conditions for entropy extremization in the bulk, and the BY tensor emerges as the canonical momentum associated with the normals on the boundary.

Before concluding this section, it is worth emphasizing that the null case exhibits a structure closely analogous to the timelike case. In particular, one can show that $\nabla_i t^i_{~j(\text{null})}= \frac{1}{4\pi}R_{ij}l^i$. Therefore, in vacuum GR with the solutions $\Lambda'=0$ one has $R_{ij}=0$, and hence the quantity $t^i_{~j(\text{null})}$ is covariantly conserved on-shell. 
Different projections of $\nabla_i t^i_{~j(\text{null})}$ lead to different physical interpretations. In particular, the contractions $l^j\nabla_i t^i_{~j(\text{null})}$, $q^j_{~a}\nabla_i t^i_{~j(\text{null})}$ and $k^j\nabla_i t^i_{~j(\text{null})}$ respectively reproduce the null Raychaudhuri equation \cite{Gourgoulhon:2005ng, Poisson:2009pwt}, the Damour--Navier--Stokes equation \cite{DAMOUR,Dey:2022fll}, and the thermodynamic identity associated with null surfaces \cite{Chakraborty:2015aja,Dey:2020tkj,Dey:2021rke,Dey:2022qqp}. Earlier, those were obtained from different projections of $R_{ab}l^b$. This further strengthens the interpretation of this $t^i_{~j(\text{null})}$ as the fundamental momentum density associated with the null hypersurface and exhibits an emergent gravity perspective in these relations.
 
 Another interesting observation is as follows. If there exists a covariant derivative $\mathcal{D}_a$ compatible to $\Pi^a_{~b}$ (though, this is a bit ad-hoc, as mentioned earlier), then one can show that $\mathcal{D}_a T^a_{~b(\text{null})} = \frac{1}{8\pi}R_{ac}l^c\Pi^a_{~b}$ \cite{Bhattacharya:2024fbz}. Hence we have $\mathcal{D}_a T^a_{~b(\text{null})} = \frac{1}{2}\Pi^a_{~b}\nabla_c t^c_{~a(\text{null})}$. This again shows that whenever the momentum conjugate to $l^a$ satisfies $\nabla_a t^a_{b~(\text{null})}\propto l_b$, one has the conservation of $T^a_{~b(\text{null})}$  within its respective manifolds.

 More importantly, it must be noticed that we have more information in $ t^i_{~j(\text{null})}$ than $T^i_{~j(\text{null})}$. In particular, it can be noted that the thermodynamic first law can be obtained from a suitable projection of $\nabla_i t^i_{~j(\text{null})}$, namely through the contraction $k^j\nabla_i t^i_{~j(\text{null})}$. However, the same thermodynamic relation cannot be recovered from $\mathcal{D}_a T^a_{~b(\text{null})}$. This issue was explicitly analyzed in \cite{Bhattacharya:2024fbz}, where it was shown that the relevant projection identically vanishes in the projected formalism. Therefore, while $T^i_{~j(\text{null})}$ captures part of the geometric information of the null surface, the full momentum tensor $t^i_{~j(\text{null})}$ retains additional dynamical and thermodynamic content which is lost upon projection.

Further, since the basis of the spacetime is $q_{ab}$ (the induced metric of the cross-section of the null surface), the null generator $l^a$ and the auxiliary null vector $k^a$, we expect that $T^a_{~b\rm{(null)}}$ should carry the individual properties of them. We will show below that the projected part of $T^a_{~b\rm{(null)}}$ on the cross-section belongs to the symmetric portion. In fact, we will be able to identify this as the conjugate momentum $\mathcal{P}^{(\textbf{q})}_{ab}$ corresponding to $q_{ab}$ in the variation of the Einstein-Hilbert action (see \ref{nullmoms}). This clarifies the symmetric nature of our construction. On the other hand $T^a_{~b\rm{(null)}} k_a$ is interestingly related to $\mathcal{P}^{(\textbf{l})}_a$.  Denote the coordinates on the null surface $\mathcal{H}$ as $x^a=(u,x^A)$, where the first coordinate corresponds to parametrizing the null generators of $\mathcal{H}$, given by $l^a = \frac{d x^a}{du}$, and $x^A$ are the coordinates on the spatial surface.  Then we have
\begin{align}
    T^A_{~~u}= T^a_{~b\rm{(null)}}l^b q^A_{~a}=& \frac{1}{8\pi}\Big[\theta^a_b+l^a\omega_b-\Pi^a_{~b}(\theta+\kappa)\Big]l^b q^A_{~a}\\
    =&\frac{1}{8\pi}\Big[l^a(\kappa)-l^a(\theta+\kappa)\Big] q^A_{~a} = 0~.
\end{align}
Next one finds
\begin{align}
    T^a_{~A}= T^a_{~b\rm{(null)}} q^b_{~A}=& \frac{1}{8\pi}\Big[\theta^a_b+l^a\omega_b-\Pi^a_{~b}(\theta+\kappa)\Big] q^b_{~A}~.
\end{align}
On using $q^a_{~b} =  \Pi^a_{~b}+l^a k_b$, in the above equation, we have
\begin{align}
    T^a_{~A}=\frac{1}{8\pi}\Big[\theta^a_{~A}+l^a q^b_{~A}\omega_b-q^a_{~A}(\theta+\kappa)\Big]~.
\end{align}
Further, we have 
\begin{align}
    T^u_{~A}=\frac{1}{8\pi}\Big[\theta^u_{~A}+l^u q^b_{~A}\omega_b-q^u_{~A}(\theta+\kappa)\Big]~,
\end{align}
which is in general non-zero.  
The projection of the null-BY tensor on the spatial surface is given by 
\begin{align}
    \Tilde{T}_{cd} = T^a_{~b\rm{(null)}}q_{ac}q^b_{~d}
    = \frac{1}{8\pi}\Big[\theta_{cd}-q_{cd}(\theta+\kappa) \Big]~, 
\end{align}
which is exactly $2\mathcal{P}^{\textbf{(q)}}_{ab}$, the conjugate momentum for the metric on 2-spatial surface $q_{ab}$. So one has $\Tilde{T}_{AB} = \frac{1}{8\pi}\Big[\theta_{AB}-q_{AB}(\theta+\kappa) \Big]$, which is symmetric.
Further, we find that 
\begin{equation}
 k_a T^a_{~b\rm{(null)}} = -\frac{1}{8\pi}\Big[\omega_b + k_b (\theta+\kappa)\Big]~,   
\end{equation}
which is $(-\mathcal{P}^{(\textbf{l})}_a)$, given in \ref{nullmoms}.

Finally, we mention that for a timelike or spacelike case, the induced metric and the normal to these hypersurfaces are enough to describe the full manifold. However, for the null case, the degenerate metric $q_{ab}$ and the null generators $l^a$ are not sufficient to foliate the full manifold, we need an auxiliary null direction, given by $k^a$.  Therefore, like $\theta_{ab}$, the second fundamental form (containing the traceless part, known as shear, and the trace part, called expansion) corresponding to $l^a$, one can associate a similar second fundamental form for $k^a$. The latter one is given by (see pages 74-75 of \cite{Gourgoulhon:2005ng})
\begin{equation}
\Xi_{ab} = \frac{1}{2}q_{ab}\theta_{(k)} + \sigma_{(k)ab}~.   
\end{equation}
Since $T^a_{~b\rm{(null)}}$ is the projected part on the null surface of $t^i_{~j\rm{(null)}}$, we expect that its covariant derivative with respect to the connection of the null hypersurface does not contain the information of $\Xi_{ab}$. This has already been mentioned in \cite{Bhattacharya:2024fbz} and also in the previous discussion. 
However it has been mentioned above that the conjugate momentum $t^i_{~j(\text{null})}$ corresponding to $l^a$ contains the thermodynamic interpretation through the quantity $k^j\nabla_i t^i_{~j(\text{null})}$. In fact such an interpretation is related to the evolution equation corresponding to $\Xi_{ab}$ (see \cite{Dey:2020tkj}). Therefore in our formalism $k^j\nabla_i t^i_{~j(\text{null})}$ bears the information of $\Xi_{ab}$.

Let us now present the relevant part which relates $\Xi_{ab}$ with $t^i_{~j(\text{null})}$. 
Following the discussion in Section-IV and Appendix-A of \cite{Dey:2020tkj}, we find
\begin{align}
    q^{ij}\,\mathcal{L}_{\ell}\Xi_{ij} = {}^{(2)}D_a \Omega^a + \Omega_a \Omega^a - \frac{1}{2}\,{}^{(2)}R + \frac{1}{2} q^{ij}R_{ij}
-\left(\kappa+\frac{\theta_{(l)}}{2}\right)\theta_{(k)}
-\frac{1}{2}\theta_{(l)}\theta_{(k)}
+2\theta_{ab}\Xi^{ab}~.
\end{align}
The use of $q^{ij}R_{ij}= R + 2R_{ij}l^i k^j$ yields
\begin{align}
    q^{ij}\,\mathcal{L}_{\ell}\Xi_{ij} = {}^{(2)}D_a \Omega^a + \Omega_a \Omega^a - \frac{1}{2}\,{}^{(2)}R + \frac{1}{2}R + R_{ij}l^i k^j
-\left(\kappa+\frac{\theta_{(l)}}{2}\right)\theta_{(k)}
-\frac{1}{2}\theta_{(l)}\theta_{(k)}
+2\theta_{ab}\Xi^{ab}~.
\end{align}
Identifying, $R_{ij}l^ik^j = 4\pi~k^j\nabla_i t^i_{~j(\text{null})}$, one finds
\begin{align}
  k^j\nabla_i t^i_{~j(\text{null})}  =-\frac{1}{4\pi} \Big[- q^{ij}\,\mathcal{L}_{\ell}\Xi_{ij} + {}^{(2)}D_a \Omega^a + \Omega_a \Omega^a - \frac{1}{2}\,{}^{(2)}R + \frac{1}{2}R 
-\left(\kappa+\frac{\theta_{(l)}}{2}\right)\theta_{(k)}
-\frac{1}{2}\theta_{(l)}\theta_{(k)}
+2\theta_{ab}\Xi^{ab}\Big]~.
\end{align}
This shows that our formalism contains the information about $\Xi_{ab}$ as well. Further the right-hand side in general does not vanish strictly on the null hypersurface as this is not restricted to this surface only. However, as mentioned above, such vanishes upon use of the on-shell condition when considered for the full-manifold.

\section{Scalar-tensor theory} \label{SECSTGRAV}
Having successfully applied our entropy-based formalism to both timelike and null hypersurfaces in general relativity, we now extend the analysis to scalar-tensor (ST) gravity in order to verify the robustness and generality of our methodology. This is a natural testbed, since higher-order $f(R)$ gravity can be recast as a special case of ST gravity \cite{Nojiri:2006ri}. Demonstrating that the Brown-York tensor can be consistently obtained within this broader class of theories further strengthens the unifying nature of our approach.

In what follows, we first consider timelike hypersurfaces in scalar-tensor gravity, closely paralleling the GR analysis. The extension to null surfaces will then follow naturally.
\subsection{Timelike Surface}
In scalar–tensor gravity, the entropy acquires modifications due to the non-minimal coupling between the Ricci scalar and the scalar field. It is well known that scalar-tensor theories can be described in two conformally related frames: the Jordan frame and the Einstein frame (see \cite{Faraoni:1999hp,Faraoni:1998qx,Faraoni:2010yi,Banerjee:2016lco,Pandey:2016unk,Bhattacharya:2017pqc,Bhattacharya:2018xlq,Bhattacharya:2020wdl,Dey:2021rke,Bhattacharya:2022mnb,Bhattacharya:2024vbx}). In the Einstein frame, the scalar field is minimally coupled and can be treated analogously to an external matter field. By contrast, in the Jordan frame, the scalar field is non-minimally coupled, so that gravity is mediated jointly by the metric and the scalar field. Since the Einstein frame closely resembles GR, our primary focus will be on the Jordan frame.

To apply our entropy-based formalism in the Jordan frame, we first postulate the form of the entropy functional. The key input is that the entropy in the two frames must be equivalent \cite{Faraoni:1999hp,Faraoni:1998qx,Bhattacharya:2017pqc,Bhattacharya:2018xlq,Bhattacharya:2022mnb}. This fact allows us to infer the Jordan-frame entropy density by starting from the Einstein-frame expression and performing the conformal transformation. Due to its close resemblance to GR, in the Einstein frame, the entropy density can be given as

\begin{align}
    \t S_{\rm E}[\t s]=\frac{1}{16\pi}\int_{\V}d^4x\tmg ~\t s_{\rm E}[\t s] ; ~~~~~~~~~~~\textrm{where}~~~~~~~~\t s_{\rm E}[\t s]=-\Big(4\t P^{cd}_{ab}\t \na_c\t s^a\t \na_d\t s^b-\t\T_{ab}(\mathbf{\t\phi,\t \psi})\t s^a\t s^b\Big)~.\label{entfunctionaleinstein}
\end{align}
Here $\tilde{\T}_{ab}(\tilde{\phi}, \tilde{\psi})$ is a symmetric rank-2 tensor, which will later be identified with the total energy–momentum tensor of the scalar field $\tilde{\phi}$ together with the additional matter fields $\tilde{\psi}$.
In the Jordan frame, we postulate an entropy functional of the form

\begin{align}
     S_{\rm J}[ s]=\frac{1}{16\pi}\int_{\V}d^4x\mg ~ s_{\rm J}[s] ;
\end{align}
Since the entropy in the two frames must be equivalent, the consistency of the formalism require $\t S_{\rm E}[\t s]=S_{\rm J}[ s]$. 

Moreover, the two frames are related by the conformal transformations provided by 
\begin{equation}
g_{ab} \to \t{g}_{ab} = \phi g_{ab},
\end{equation}
along with
\begin{equation}
d\tilde{\phi} = \sqrt{\frac{2\omega(\phi)+3}{16\pi}} \,\frac{d\phi}{\phi}.
\end{equation}
Using $\tmg=\phi^2 \mg$, we obtain
\begin{align}
    s_{\rm J}[s]=\phi^2\t s_{\rm E}[\t s]=-4\phi^2\t P^{cd}_{ab}\t \na_c\t s^a\t \na_d\t s^b+\phi^2\t\T_{ab}(\t\phi,\t\psi)\t s^a\t s^b~.\label{EE1}
\end{align}
To express this in terms of Jordan-frame variables, we note that normal vector transforms under the conformal rescaling as $\t s^a=s^a/\sqrt{\phi}$, implying the change of its covariant derivative as
\begin{align}
    \t\na_c\t s^a=\frac{1}{\sqrt{\phi}}\na_cs^a+\frac{1}{2\phi^{\frac{3}{2}}}\d^a_cs^i\na_i\phi-\frac{1}{2\phi^{\frac{3}{2}}}s_c\na^a\phi~.
\end{align}
Substituting this relation into the expression for $s_{\rm J}[s]$, and simplifying, we finally obtain
\begin{align}
    s_{\rm J}[s]=-\Big(4\phi P^{cd}_{ab}\na_cs^a\na_ds^b-\T_{ab}(\phi,\psi)s^as^b\Big)-4(\na_a s^a)(s^b\na_b\phi)-\frac{3}{\phi}s^as^b(\na_a\phi)(\na_b\phi) -2 (s_a \na^b\phi)(\na_b s^a)~,
    \label{BB11}
\end{align}
where $\T_{ab}(\phi,\psi)=\phi \t\T_{ab}(\t\phi,\t\psi)$.  Thus, taking the variation of the action functional $S_{\rm J}[ s]$ due to $s^a\rightarrow s^a+\d s^a$ yields
\begin{align}
    \d S_{\rm J}[ s]=-\frac{1}{8\pi}\int_{\V}d^4x\mg~\Big[\Big(4\phi P^{cd}_{ab}(\na_c\d s^a)(\na_ds^b)- \T_{ab}(\phi,\psi)\d s^as^b\Big)+2(\na_a \d s^a)(s^b\na_b\phi)
    \no 
    \\
    +2(\na_b s^b)(\na_a\phi)\d s^a+\frac{3}{\phi}s^b(\na_a\phi)(\na_b\phi)\d s^a+ \delta s^a(\na_b s_a)\na^b\phi+(s_a\na^b\phi)(\na_b \delta s^a)\Big]~.
\label{VarJor}
\end{align}
Integrating by parts, one can obtain the bulk part and the boundary term, which is given as
\begin{align}
    \d S_{\rm J}[ s]=\frac{1}{8\pi}\int_{\V}d^4x\mg~\Big[4\phi P^{cd}_{ab}(\na_c\na_ds^b)+2s^b\na_a\na_b\phi-\frac{3}{\phi}s^b(\na_a\phi)(\na_b\phi)
    \no\\
    + \Box\phi\, g_{ab} s^b + \T_{ab}(\phi,\psi)s^b\Big]\d s^a+\int_{\p\V}d\Sigma_i t^i_{~j\rm{(Jor)}}\d s^j~, \label{finvarSJ}
\end{align}
where the canonically conjugate surface term can be identified as 
\begin{align}
    t^i_{~j\rm{(Jor)}}=\frac{\p s_{\rm J}[s]}{\p(\na_is^j)}=-\frac{1}{16\pi}\Big[8\phi P^{id}_{jb}\na_ds^b+4\d^i_j ~s^b\na_b\phi + 2s_j\na^i \phi\Big]\no 
    \\
    =\frac{1}{16\pi}\Big[4\phi\big(\na_js^i-\delta^i_j\na_as^a\big)-4\d^i_j ~s^b\na_b\phi -2s_j\na^i \phi\Big]~.
    \label{BB12}
\end{align}
Again, projecting onto the hypersurface, with appropriate normalisation, yields
\begin{align}
    T^{a\rm (BY)}_{b}(\boldsymbol{\gamma},\phi)=-\frac{1}{2}\g^a_i\g^j_bt^i_{~j\rm{(Jor)}}=\frac{1}{8\pi}\Big[\phi(K^a_b-\g^a_b K)+\g^a_bs^i\na_i\phi\Big]~,
    \label{BYTL}
\end{align}
which is the BY tensor for a timelike surface in the Jordan frame of scalar-tensor gravity \cite{Bhattacharya:2023ycc,Creighton:1995au,Bose:1998yp,Cote:2019fkf}. Furthermore, as was the case in general relativity, the equation of motion can be obtained from the bulk part, which has been obtained in the later part.

\subsection{Null Surface}
The above analysis for the timelike surface can be extended to the null surface, simply by replacing the timelike normals with the null ones. Again from the argument about the equivalence of the entropy, the entropy functional for the null surface can be postulated as
\begin{align}
     S_{\rm J}[ l]=\frac{1}{16\pi}\int_{\V}d^4x\mg ~ s_{\rm J}[l] ;
\end{align}
where the entropy density functional can be defined as
\begin{align}
    s_{\rm J}[l]=-\Big(4\phi P^{cd}_{ab}\na_cl^a\na_dl^b-\T'_{ab}(\phi,\psi)l^al^b\Big)-4(\na_a l^a)(l^b\na_b\phi)-\frac{3}{\phi}l^al^b(\na_a\phi)(\na_b\phi)-2 (l_a \na^b\phi)(\na_b l^a)~.
    \label{lagnull}
\end{align}
Variation of the action for $l^a\to l^a+\delta l^a$ is given by
\begin{align}
    \d S_{\rm J}[ l]=\frac{1}{8\pi}\int_{\V}d^4x\mg~\Big[4\phi P^{cd}_{ab}(\na_c\na_dl^b)+2l^b\na_a\na_b\phi-\frac{3}{\phi}l^b(\na_a\phi)(\na_b\phi)
    \no\\
    + \T'_{ab}(\phi,\psi)l^b + \Box\phi\, g_{ab} l^b\Big]\d l^a +\int_{\p\V}d\Sigma_i t^{i \rm (null)}_{~j\rm{(Jor)}}\d l^j~, \label{finVarSJ}
\end{align}
Here we defined
\begin{align}
    t^{i \rm (null)}_{~j\rm{(Jor)}}=\frac{\p s_{\rm J}[l]}{\p(\na_il^j)}
=-\frac{1}{16\pi}\Big[8\phi P^{id}_{jb}\na_dl^b+4\d^i_j ~l^b\na_b\phi +2 l_j\na^i \phi\Big]
\no
\\
=\frac{1}{16\pi}\Big[4\phi\big(\na_jl^i-\delta^i_j\na_al^a\big)-4\d^i_j ~l^b\na_b\phi -2 l_j\na^i \phi\Big]
    \no 
    \\
    =\frac{1}{16\pi}\Big[4\phi\Big(\theta^i_j +l^i\omega_j-l_jk^m\na_ml^i-\d^i_j(\theta+\kappa) \Big)-4\d^i_j ~l^b\na_b\phi - 2 l_j\na^i \phi\Big]~.
    \label{BB13}
\end{align}
Then the Brown-York tensor turns out to be
\begin{align}
    T^{a\rm (null)}_{~~b}(\Pi,\phi)=\frac{1}{2}\Pi^a_{~i}\Pi^j_{~b}t^i_{j\rm{(null)}}=\frac{1}{8\pi}\Big[\phi\Big(\theta^a_b+l^a\omega_b-\Pi^a_{~b}(\theta+\kappa)\Big)-\Pi^a_{~b}~l^i\na_i\phi\Big]~.
    \label{STBY2}
\end{align}
Having successfully obtained the Brown-York tensor from the entropy functional and thereby extending the formalism beyond Einstein gravity, we now turn to the derivation of the bulk gravitational field equations in scalar–tensor gravity. Unlike the case of general relativity, where Padmanabhan’s entropy functional has been extensively explored, its extension to scalar-tensor theories and the associated variational structure have received comparatively little attention in the literature. In particular, it is important to understand how the additional scalar degree of freedom modifies the thermodynamic variational principle and alters the conservation properties of the corresponding boundary momentum tensor. In the following, we show that the same entropy-based framework naturally reproduces the bulk field equations of scalar-tensor gravity while simultaneously providing a consistent interpretation of the generalized Brown-York tensor. Though the following calculations are done for spacelike normals, they are valid for null normals as well.

Note that, for the scalar-tensor theory, the conjugate momenta (\ref{BB12} and \ref{BB13}) are not covariantly conserved by themselves. This non-conservation has a natural interpretation in terms of momentum exchange between the hypersurface-normal sector and the scalar sector, which are coupled through the scalar-tensor interaction. In other words, the normal vector and the scalar field should not be regarded as independently conserved sectors; rather, their individual momentum balances are modified by their mutual coupling. We will see below that this exchange is also reflected in the corresponding boundary dynamics.

As shown in \ref{App2} and \ref{appnull}, the Brown--York counterparts obtained from these momenta are likewise not conserved individually on their respective hypersurfaces. The situation is somewhat more subtle than what one might infer from the standard metric-based construction. In order to make the origin of this non-conservation transparent, we establish, in analogy with the GR case, a direct relation between the intrinsic divergence of the Brown--York tensor and the tangential projection of the covariant divergence of the momentum conjugate to the hypersurface normal. For the timelike and null hypersurfaces, respectively, we find
\begin{align}
D_i T^{i\rm{(BY)}}_b = -\frac{1}{2}\gamma^j_{~b}\na_i t^i_{~j\rm{(Jor)}} - \frac{1}{16\pi} \na^i\phi (\na_i s_b)~,
\label{appeq12}
\end{align}
and 
\begin{align}
\mathcal{D}_i  T^{i\rm (null)}_{~~b} = \frac{1}{2} \Pi^j{}_b \nabla_i  t^{i \rm (null)}_{~j\rm{(Jor)}} + \frac{1}{16\pi} \left[ \nabla^i \phi \, \theta_{ib} -(l^a \nabla_a \phi) \Pi^i{}_b (k^m \nabla_m l_i) \right]~.
\label{Ref1}
\end{align}
The detailed derivations are provided in \ref{AppRef}. These relations make explicit that the non-conservation of the Brown-York tensor is not an isolated failure of the boundary momentum balance. Rather, in addition to the projected divergence of the normal-vector momentum, there are explicit contributions involving the scalar field and its derivatives. These terms encode the exchange of momentum between the gravitational boundary degrees of freedom associated with the hypersurface normal and the scalar sector.

An important consistency check is obtained by setting $\phi=1$. In this limit, all explicit scalar contributions vanish, and Eqs.~\eqref{appeq12} and \eqref{Ref1} reduce to the corresponding relations obtained in general relativity. Thus, the GR conservation relation is recovered when the scalar-tensor coupling is switched off.
Further this approach helps to identify the alternative perspective of the Brown-York tensor – this is the projected part of the conjugate momentum corresponding to the normal to the hypersurface in the entropy functional action.

\subsection{Obtaining bulk gravitational equation of motion}
Our main target was to identify the Brown-York tensor, which we obtained successfully. For the purpose of completeness, here we show how the equations of motion can be obtained within this formalism.
\subsubsection{Einstein frame}
We begin in the Einstein frame, where the entropy density is given by \ref{entfunctionaleinstein}. Variation with respect to the vector field $\t s^a$ yields the bulk equation (similar to Einstein's gravity):
\begin{align}
    \tilde{R}_{ab} = \frac{1}{2}\bar{\tilde{\mathcal{T}}}_{ab}(\tilde{\phi},\tilde{\psi}), \label{eomein}
\end{align}
where $ \bar{\tilde{\mathcal{T}}}_{ab}(\tilde{\phi},\tilde{\psi})=\tilde{\mathcal{T}}_{ab}(\tilde{\phi},\tilde{\psi})+\t\Lambda \t g_{ab}$ with $\t\Lambda$ being the Lagrange multiplier.\\
This further yields 
\begin{align}
    \tilde{G}_{ab} = \frac{1}{2}\bar{\tilde{\mathcal{T}}}_{ab}(\tilde{\phi},\tilde{\psi})-\frac{1}{4}\t g_{ab} \bar{\tilde{\mathcal{T}}}(\tilde{\phi},\tilde{\psi})=8\pi \t{\mathfrak{T}}_{ab}(\tilde{\phi},\tilde{\psi})~. \label{EOMEIN}
\end{align}
At this stage, the tensor $\t{\mathfrak{T}}_{ab}(\tilde{\phi},\tilde{\psi})$ is not fixed by the variational principle and remains arbitrary within the entropic formulation. However, since the scalar field is minimally coupled in the Einstein frame, this tensor can be naturally identified as the sum of matter and scalar contributions, along with additional term arising from the Lagrange multiplier i.e.
\begin{align}
    \t{\mathfrak{T}}_{ab}(\tilde{\phi},\tilde{\psi})=\t T_{ab}^{\rm (mat)}(\psi)+\t T_{ab}^{(\t\phi)}-\frac{1}{8\pi}\t\Lambda(\tilde{\phi},\tilde{\psi})\t g_{ab}~, \label{tabphipsitil}
\end{align}
where $\t\Lambda(\tilde{\phi},\tilde{\psi})$ is the contribution from Lagrange multiplie, $\t T_{ab}^{\rm (mat)}(\psi)$ is the energy-momentum tensor of the external matter and $\t T_{ab}^{(\t\phi)}$ is the energy-momentum tensor of the scalar field $\t\phi$ i.e.
\begin{align}
    \t T_{ab}^{(\t\phi)}= (\t\na_a\t\phi)(\t\na_b\t\phi)-\frac{1}{2}\t g_{ab}(\t\na^i\t\phi)(\t\na_i\t\phi)-\t g_{ab}U(\t\phi)~.
\label{EMTPHITIL}
\end{align}
Its conservation relation can be obtained as
\begin{align}
    \t\na_a \t T^{ab}(\t\phi)=\left(\t\square\t\phi-\frac{dU}{d\t\phi}\right)(\t\na^b\t\phi)=\t E_{(\t\phi)}(\t\na^b\t\phi)~,
\end{align}
which implies that $T_{ab}(\t\phi)$ is conserved onshell.
Furthermore, as the energy-momentum tensor of the external field is conserved, it implies that it $\t\Lambda(\tilde{\phi},\tilde{\psi})$ has to either vanish or be constant to remain consistent with the generalized Bianchi identity (in the latter case, it will only provide a shift on $U(\t\phi)$ ).
It ensures the consistency of the construction and further justifies the identification of $\t{\mathfrak{T}}_{ab}(\tilde{\phi},\tilde{\psi})$ as the effective total energy-momentum tensor in the Einstein frame, and with that identification, \ref{EOMEIN} can be identified as the gravitational equation of motion in the Einstein frame. Finally, the corresponding equation of motion can be obtained as
\begin{align}
    \tilde{G}_{ab} + \tilde{\Lambda}\tilde{g}_{ab}= 8\pi\Big[(\t\na_a\t\phi)(\t\na_b\t\phi)-\frac{1}{2}\t g_{ab}(\t\na^i\t\phi)(\t\na_i\t\phi)-\t g_{ab}U(\t\phi)+\t T_{ab}^{\rm (mat)}(\psi)\Big]~.
\end{align}
\subsubsection{Jordan frame}
Upon using the extremization of the entropy functional and fixing the normal on the boundary in the Jordan frame, from \ref{finvarSJ}, we have 

\begin{align}
    4\phi P^{cd}_{ab}(\na_c\na_ds^b)+2s^b\na_a\na_b\phi-\frac{3}{\phi}s^b(\na_a\phi)(\na_b\phi)
    + \T_{ab}(\phi,\psi)s^b +\Box \phi\, g_{ab} s^b =0~.
\end{align}
Using the commutation relation of the covariant derivative, one can further obtain
\begin{align}
    -2 \phi R_{ab}+2\na_a\na_b\phi-\frac{3}{\phi}(\na_a\phi)(\na_b\phi) + \Box \phi\, g_{ab} + \T_{ab}(\phi,\psi)=0~.
\label{phirab}
\end{align}
From \ref{eomein} one can obtain on-shell:
\begin{align}
    \t{{\mathcal{T}}}_{ab}(\t{\phi},\t{\psi}) = \t R\, \t g_{ab} + 16\pi \t T^{(mat)}_{ab} + 16\pi \big[(\t\na_a\t\phi)(\t\na_b\t\phi)-\frac{1}{2}\t g_{ab}(\t\na^i\t\phi)(\t\na_i\t\phi)-\t g_{ab}U(\t\phi) \big] -3\t\Lambda(\tilde{\phi},\tilde{\psi})\t g_{ab}~.
    \label{reqt}
\end{align}
Upon conformal transformation, this boils down to
\begin{align}
   \t{{\mathcal{T}}}_{ab}(\t{\phi},\t{\psi}) = \Big( R\,g_{ab} + \frac{3}{2\phi^2}\na^i\phi \na_i\phi\, g_{ab} - \frac{3}{\phi}\square\phi\, g_{ab} \Big) + 16\pi \t T^{(mat)}_{ab} + \Big[ \Big(\frac{3+2\omega}{ \phi^2}\Big)\na_a \phi \na_b \phi\no \\
 - \frac{1}{2}(g_{ab})\Big(\frac{3+2\omega}{\phi^2}\Big)\na_i \phi \na^i\phi - 16 \pi \phi g_{ab} U(\t \phi)\Big] -3\phi\,\t\Lambda(\tilde{\phi},\tilde{\psi})g_{ab}~.
\end{align}
Substituting these relations in \ref{phirab}, one obtains 
\begin{align}
\phi G_{ab} =  8\pi T^{\rm (mat)}_{ab} + \frac{\omega}{\phi}\Big(\na_a \phi \na_b \phi - \frac{1}{2}g_{ab}\na^i\phi \na_i \phi\Big) + \na_a\na_b\phi - g_{ab} \square\phi - \frac{V(\phi)}{2}g_{ab}~, \label{idenjor1}
\end{align}
where $V(\phi) = 16 \pi \phi^2 U(\tilde{\phi}) + 3 \phi^2 \tilde{\Lambda}$. 
This is the standard equation of motion in the Jordan frame.

The same steps apply to the null case as well, leading to the same equations of motion.

\section{Discussions and conclusions} \label{concl}

In this work, we have developed a unified variational framework to understand the origin and interpretation of the Brown-York (BY) tensor across both timelike and null hypersurfaces, in general relativity and beyond. By drawing inspiration from the entropy-based formulation of gravity by Padmanabhan \cite{Padmanabhan:2004kf, Padmanabhan:2007en}, we have shown that the BY tensor admits a natural interpretation as the projection of the momentum density conjugate to geometric variables associated with hypersurface normals, rather than as a conventional energy-momentum tensor.
This perspective provides a coherent explanation for several longstanding subtleties. In particular, the difficulty in defining the BY tensor on null hypersurfaces is traced back to the degeneracy of the induced metric and the limitations of the standard Hamilton-Jacobi prescription. Within our framework, however, the construction remains well-defined and does not rely on ad hoc prescriptions even for the null surfaces.

In this connection, we would like to emphasize an important and well-known aspect of variational principles. At the level of the bulk equations of motion, an action is defined only up to an additive constant and, more generally, up to a total derivative. Such modifications do not alter the bulk Euler--Lagrange equations, provided the corresponding boundary conditions are appropriately specified. However, when the boundary contribution is itself physically relevant, as in the present construction, this freedom requires further qualification. In particular, the addition of a total derivative modifies the boundary potential (or symplectic potential) and can consequently change the canonical momentum associated with the boundary data. Thus, although two Lagrangians may be equivalent from the viewpoint of their bulk equations of motion, they need not lead to identical boundary canonical momenta.

This distinction is particularly important in the present context because the functional employed here is not introduced merely as an auxiliary variational device for reproducing the gravitational equations. Following Padmanabhan's emergent-gravity paradigm, the entropy functional acquires a direct thermodynamic interpretation on shell as a spacetime entropy functional; see, in particular, Sec. III of Ref.~\cite{Padmanabhan:2007en}. Consequently, an arbitrary modification by a total divergence cannot automatically be regarded as physically innocuous. Such a modification changes the boundary contribution of the functional and may therefore alter its thermodynamic interpretation as well as the associated boundary canonical momentum. At the same time, this does not imply that every total derivative is forbidden: boundary modifications that preserve the relevant thermodynamic interpretation, boundary conditions, and variational structure may still be admissible. The physically meaningful question is therefore not whether the entropy functional is mathematically unique under arbitrary integrations by parts, but whether the admissible representative is uniquely fixed once the thermodynamic and variational requirements are imposed.

For this reason, throughout this work we employ the canonical entropy functional originally proposed by Padmanabhan. Its form is not chosen arbitrarily, but is selected within the relevant class of functionals by a combination of physical requirements, including translational invariance, the quadratic dependence on the derivatives of the deformation field, the correct thermodynamic interpretation of the on-shell functional, and the requirement that its extremization reproduce the gravitational field equations. In the scalar--tensor case, we do not introduce an independent entropy functional by an unrelated ansatz; rather, the corresponding Jordan-frame functional is obtained by transforming the canonical Einstein-frame functional under the appropriate conformal transformation and scalar-field redefinition. This provides a definite representative within the class of variationally equivalent functionals.

Within this physically distinguished choice, the canonical momentum conjugate to the hypersurface normal is well defined, and its appropriate tangential projection yields the Brown--York tensor. We stress, however, that the resulting boundary momentum can in principle be sensitive to admissible boundary modifications of the underlying functional. This is not a peculiarity of the present entropy-based formulation. An analogous issue is already present in the conventional variational construction of the Brown--York tensor, where the boundary term added to the gravitational action is essential for obtaining a well-defined variational principle and determines the corresponding boundary response. In this sense, the boundary-momentum ambiguity encountered here is part of the general structure of gravitational theories with boundaries rather than a new ambiguity introduced by our approach.

Our claim, therefore, is not that the entropy functional is absolutely unique with respect to all possible off-shell redefinitions. Rather, we work with the physically distinguished Padmanabhan representative and show that, for this representative, the Brown--York tensor follows as the appropriate projection of the entropy-functional polymomentum. The same choice is then consistently carried through the timelike, null, and scalar--tensor constructions developed in this work.

A key result of our analysis is the identification of the structural origin of the conservation of the Brown-York (BY) tensor. In general relativity, in absence of matter the covariant derivative of the conjugate momentum corresponding to the hypersurface normal is proportional to the normal itself when on shell condition is considered. Further, the projected part of the covariant derivative of the canonical momentum is directly related to the boundary covariant derivative of the BY tensor. Therefore, it naturally yields the conservation of the BY tensor. This property, however, is not generic to extensions of general relativity. In scalar-tensor theories, the coupling between the scalar and gravitational sectors breaks the only-linear dependence of covariant derivative of conjugate momentum on the hypersurface normal, leading to a non-conservation of the corresponding momentum. Further related BY counterparts are not conserved on the boundary. This non-conservation is naturally interpreted as a momentum exchange between the gravitational and scalar sectors, with the scalar response providing the corresponding source in the boundary momentum balance. However, within the closed system, composed of the boundary gravitational field and the scalar field, there is no non-conservation. Thus, our framework provides a unified interpretation of both the conservation of the BY tensor in general relativity and its modified conservation law in scalar-tensor gravity.
We emphasize that the present work provides an alternative variational perspective in which the Brown-York tensor arises as the tangential projection of the entropy-functional polymomentum. In this formulation, the null structure is incorporated at the level of the variational principle itself, which is rooted in the entropy-based emergent gravity framework. 
The summary of the main features of our investigation is as follows.
\begin{itemize}
\item Our starting point is not the Einstein-Hilbert action. Instead, we adopted a different viewpoint in which gravity is regarded as an emergent theory. The dynamics of gravitational degrees of freedom is governed by the extremization of gravitational entropy, and consequently, Einstein's equation is interpreted as a thermodynamic identity as well as a fluid-type equation of motion. In this context, the concept of entropy functional is very important and provides the backbone of the emergent nature of gravity. Hence, this is a completely different approach and provides an alternative interpretation of the Brown-York tensor.
\item In most of the literature, the derivation of the Brown-York (BY) tensor is based on the stretched horizon approach. Our construction of the BY stress-tensor, which is defined on a null surface, has an important property. Note that in our construction, the spacetime is foliated by a stack of null-surfaces, and hence the tensor is applicable to all null-surfaces.
Therefore, it is natural to use this BY tensor for the null-foliation case. This particular thing is missing in the stretched horizon approach, where the null surface is a particular hypersurface on which the spacelike normal to the timelike surfaces, foliating the spacetime, becomes null (like the black hole event horizon \cite{Parikh:1997ma,Agca:2024dlf}). 
\item We found that the BY tensor, in the context of entropic function, is connected to the conjugate momentum corresponding to the generator $l^a$ of the null surface. This is a new finding and provides an important aspect within the pre-existing idea of the emergent nature of gravity.
\end{itemize}

More broadly, our results highlight a common underlying structure linking bulk gravitational dynamics and boundary-defined quantities. The entropy-based variational principle not only reproduces the equations of motion but also simultaneously encodes the boundary momentum, offering a deeper interpretation of quasi-local gravitational quantities. More importantly, it has been shown that the results can be extended to modified theories of gravity as well, implying the robustness of the formalism. We believe that our framework provides useful insights that may aid further developments in this area.

\section*{Acknowledgement}
The research of KB is supported by the New Faculty Seed Grant (NFSG) of BITS Pilani Dubai Campus. The work of BR is supported by the University Grants Commission (UGC), Government of India, under the scheme Junior Research Fellowship (JRF).
The research of BRM is supported by the Science and Engineering Research Board (SERB), Department of Science $\&$ Technology (DST), Government of India, under the scheme Core Research Grant (File no. CRG/2020/000616).

\appendix
\section*{\Huge{Appendices}}
\section{Boundary divergence of BY tensor in ST gravity} 
\setcounter{equation}{0}
\renewcommand{\theequation}{A.\arabic{equation}}
\subsection{Timelike surface} \label{App2}
The BY tensor in the Jordan frame for timelike surface is given by \ref{STBY1}.
Let $D_a$ be the covariant derivative which is metric compatible on the timelike surface. On taking the covariant derivative of the above equation, we have
\begin{align}
    D_a T^{ab\rm (BY)}(\boldsymbol{\gamma},\phi)=\frac{1}{8\pi}\Big[D_a\phi\Big(K^{ab}-K\gamma^{ab}\Big)+ \phi D_a\Big(K^{ab}-K\gamma^{ab}\Big)+\gamma^{ab}D_a\Big(s^i\na_i\phi\Big)\Big]~.
\end{align}
Manipulating the last term in the above equation, one has 
\begin{eqnarray}
    D_a T^{ab\rm (BY)}(\boldsymbol{\gamma},\phi)&=&\frac{1}{8\pi}\Big[D_a\phi\Big(K^{ab}-K\gamma^{ab}\Big)+ \phi D_a\Big(K^{ab}-K\gamma^{ab}\Big)+\underbrace{\gamma^{ab}\Big(\na_a s^i\Big)}_{-K^{bi}}\na_i\phi + \gamma^{ab} s^i \Big(\na_a \na_i \phi \Big)\Big]
    \nonumber
    \\
    &=&\frac{1}{8\pi}\Big[-\Big(K\gamma^{ab}\Big)\na_a \phi+ \phi D_a\Big(K^{ab}-K\gamma^{ab}\Big) + \gamma^{ab} s^i \Big(\na_a \na_i \phi \Big)\Big]~.
\end{eqnarray}
Using the geometric identity $D_a\Big(K^{a}_b-K\gamma^{a}_b\Big) = -R_{ac}\gamma^a_b s^c$, we have 
\begin{align}
    D_a T^{ab\rm (BY)}(\boldsymbol{\gamma},\phi)=\frac{1}{8\pi}\gamma^{ab}\Big[-K \na_a \phi - \phi R_{ac} s^c +  s^i \na_a \na_i \phi \Big]~.
    \label{Eq102}
\end{align}
Multiplying \ref{EoMST} by $\gamma^{ab}s^c$, it yields
\begin{align}
\Big(\phi R_{ac}s^c - s^c \na_a \na_c \phi \Big)\gamma^{ab} =  8\pi T^{\rm{(mat)}}_{ac} \gamma^{ab} s^c + \frac{\omega}{\phi}\nabla_a\phi\nabla_c\phi \gamma^{ab} s^c~.
\label{geoiden}
\end{align}
Now, use of \ref{geoiden} into \ref{Eq102}, we obtain \ref{Conservation1}.

\subsection{Null surface} \label{appnull}
Below we present similar calculations in the context of the null case. 
The BY tensor in the Jordan frame for the null surface is given by \ref{STBY2}. 
Then one finds
\begin{align}
\mathcal{D}_a T^{a\rm (null)}_{~~b}(\Pi,\phi)= \frac{1}{8\pi}\Big[\mathcal{D}_a\phi\Big(\theta^a_b+l^a\omega_b-\Pi^a_{~b}(\theta+\kappa)\Big) + \phi \mathcal{D}_a\Big(\theta^a_b+l^a\omega_b-\Pi^a_{~b} (\theta+\kappa)\Big)-\Pi^a_{~b}~\mathcal{D}_a \big(l^i\na_i\phi\big)\Big]~.
\label{app1}
\end{align}
A little manipulation of the last term in the above equation yields, 
\begin{align}
\Pi^a_{~b}~\mathcal{D}_a \big(l^i\na_i\phi\big)  =  \mathcal{D}_a\phi\Big(\theta^a_b+l^a\omega_b\Big) + \Pi^c_{~b} l^i \na_c \na_i \phi~.   
\label{app3}
\end{align}
Use of \ref{app3} into \ref{app1} gives
\begin{align}
\mathcal{D}_a T^{a\rm (null)}_{~~b}(\Pi,\phi)= \frac{1}{8\pi}\Big[-\Pi^a_{~b}(\theta+\kappa)\mathcal{D}_a\phi + \phi \mathcal{D}_a\Big(\theta^a_b+l^a\omega_b-\Pi^a_{~b} (\theta+\kappa)\Big)-\Pi^a_{~b}~l^i \na_a \na_i\phi\Big]~.
\label{app4}
\end{align}
Upon using the geometric identity, $\mathcal{D}_a\Big(\theta^a_b+l^a\omega_b-\Pi^a_{~b} (\theta+\kappa)\Big) = R_{ac} \Pi^a _{~b} l^c$ (see, \cite{Bhattacharya:2024fbz}) one finds
\begin{align}
\mathcal{D}_a T^{a\rm (null)}_{~~b}(\Pi,\phi)= \frac{1}{8\pi} \Big[ \big( \phi R_{ac}-\na_a 
\na_c \phi\big)\Pi^a_{~b}l^c - \Pi^a_{~b}\big(\theta+\kappa\big)\na_a \phi\Big]~. 
\label{app5}
\end{align}
Multiplying \ref{EoMST} by $\Pi^{a}_{~b} l^c$, it yields
\begin{align}
\Big(\phi R_{ac} - \na_a \na_c \phi \Big)\Pi^{a}_{~b}l^c =  8\pi T^{\rm{(mat)}}_{ac} \Pi^{a}_{~b} l^c + \frac{\omega}{\phi}\nabla_a\phi\nabla_c\phi \Pi^{a}_{~b} l^c~.
\label{geoiden2}
\end{align}
Now, use of \ref{geoiden2} in \ref{app5}, we obtain, 
\begin{align}
    \mathcal{D}_a T^{a\rm (null)}_{~~b}(\Pi,\phi)=T_{ac}^{\rm{(mat)}}\Pi^a_{~b} l^c - \Pi^{\rm{(Null)}}_{(\phi)}\Pi^a_{~b}\mathcal{D}_a \phi~,
    \label{Conservationnull}
\end{align}
where, $\Pi^{(\phi)} =  \frac{1}{8\pi}\big(\theta +\kappa - \frac{\omega}{\phi}l^c \na_c \phi\big)$ is the conjugate quantity of $\phi$ on a null hypersurface \cite{Bhattacharya:2023ycc}. 

\section{Derivations of \ref{appeq12} and \ref{Ref1}}\label{AppRef}
\setcounter{equation}{0}
\renewcommand{\theequation}{B.\arabic{equation}}
\subsection{Timelike hypersurface} \label{App3}
The momentum conjugate to the normal field $s^a$ is given by \ref{BB12}. On taking the covariant derivative of \ref{BB12}, we have
\begin{align}
    \na_i t^i_{~j\rm{(Jor)}} &=\frac{1}{16\pi}\Big[4\na_i\phi\big(\na_js^i-\delta^i_j\na_as^a\big) + 4\phi \na_i\big(\na_js^i-\delta^i_j\na_as^a\big)-4\na_j \big(s^b\na_b\phi\big) -2\na_i\big(s_j\na^i \phi\big)\Big]
    \nonumber
    \\
    &=\frac{1}{16\pi}\Big[-4\na_j\big(\phi\,\na_i s^i\big) + 4 \na_i\big(\phi \na_js^i\big)-4\na_j \big(s^b\na_b\phi\big) -2\na_i\big(s_j\na^i \phi\big)\Big]~.
    \label{appeq3}
\end{align}
Using $\na_i s^j = -K_i^{~j} + s_i a^j$ and $\na_i s^i = -K$, the above equation reduces to, 
\begin{align}
    \na_i t^i_{~j\rm{(Jor)}}=\frac{1}{16\pi}\Big[4\na_j\big(\phi\, K\big) + 4 \na_i\big\{\phi \big(-K^i_{~j} + s_j a^i\big)\big\}-4\na_j \big(s^b\na_b\phi\big) -2\na_i\big(s_j\na^i \phi\big)\Big]~.
    \label{appeq4}
\end{align}
A little bit of rearrangement yields,
\begin{align}
    \na_i t^i_{~j\rm{(Jor)}}=\frac{1}{4\pi}\na_i\Big[\phi\big(\delta^i_j K -K^i_{~j} + s_j a^i\big)-\delta^i_j \big(s^b\na_b\phi\big)\Big] -\frac{1}{8\pi}\na_i\big(s_j\na^i \phi\big)~.
    \label{appeq5}
\end{align}
Substituting $\gamma^i_j = \delta^i_j - s^i s_j$, we have,
\begin{align}
    \na_i t^i_{~j\rm{(Jor)}}=-\frac{1}{4\pi}\na_i\Big[\phi\big(K^i_{~j} - \gamma^i_{~j} K \big)+ \gamma^i_{~j} \big(s^b\na_b\phi\big)\Big] + \frac{1}{4\pi}\na_i\Big[\underbrace{\phi K s^is_j+ \phi s_j a^i -s^i s_j s^a\na_a \phi-\frac{1}{2}\big(s_j\na^i \phi\big)}_{A^i_{~j}}\Big]~.
    \label{appeq6}
\end{align}
Identifying the first term in the bracket as the BY tensor given by \ref{BYTL}, and calling the terms within the third bracket in the last term as $A^i_{~j}$, we have
\begin{align}
    \na_i t^i_{~j\rm{(Jor)}}=-2\na_i T^{i\rm{(BY)}}_{~j}+ \frac{1}{4\pi}\na_i A^i_{~j}.
    \label{appeq7}
\end{align}
On multiplying \ref{appeq7} with $\gamma^j_{~b}$, we have
\begin{align}
-\frac{1}{2}\gamma^j_{~b}\na_i t^i_{~j\rm{(Jor)}}=\gamma^j_{~b}\na_i T^{i\rm{(BY)}}_{~j} - \frac{1}{8\pi}\gamma^j_{~b}\na_i A^i_{~j}.
\label{appeq8}
\end{align}
Let's manipulate both the terms on the RHS of the above equation separately. Upon focusing on the first term of \ref{appeq8},
\begin{align}
\nonumber
\gamma^j_{~b}\na_i T^{i\rm{(BY)}}_{~j} &= \gamma^j_{~b} \delta^m_i \delta^i_n \na_m T^{n\rm{(BY)}}_{~j}\\ 
\nonumber
&= \gamma^j_{~b} (\gamma^m_{~n} + s^m s_n) \na_m T^{n\rm{(BY)}}_j\\
\nonumber
&= \gamma^j_{~b}\gamma^m_{~n}\na_m T^{n\rm{(BY)}}_j + \gamma^j_{~b} s^m s_n
\na_m T^{n\rm{(BY)}}_{~j}\\
&= \gamma^j_{~b}\gamma^m_{~i}\gamma^i_{~n}\na_m T^{n\rm{(BY)}}_j + \gamma^j_{~b} s^m s_n \na_m T^{n\rm{(BY)}}_{~j}~;
\end{align}
which simplifies to, 
\begin{align}
\gamma^j_{~b}\na_i T^{i\rm{(BY)}}_j &= D_i T^{i\rm{(BY)}}_b + \gamma^j_{~b} s^m s_n \na_m T^{n\rm{(BY)}}_j~.
\label{appeq14}
\end{align}
Further, since the BY tensor lies on the hypersurface normal to $s^a$, we have,
\begin{align}
\nonumber
\gamma^j_{~b}\na_i T^{i\rm{(BY)}}_j &= D_i T^{i\rm{(BY)}}_b + \gamma^j_{~b} s^m s_n \na_m T^{n\rm{(BY)}}_j\\
\nonumber
&= D_i T^{i\rm{(BY)}}_b - \gamma^j_{~b} T^{n\rm{(BY)}}_j (\na_m s_n) s^m\\
&= D_i T^{i\rm{(BY)}}_b -  T^{i\rm{(BY)}}_b a_i~. 
\label{appeq13}
\end{align}
On plugging the expression for the BY tensor from \ref{BYTL} in \ref{appeq13}, we have
\begin{align}
\gamma^j_{~b}\na_i T^{i\rm{(BY)}}_j &= D_i T^{i\rm{(BY)}}_b - \frac{1}{8\pi} \Big[\phi K^i_{~b}a_i - a_b\phi K + a_b s^j\na_j \phi\Big]~.
\label{appeq9}
\end{align}

Now, let's focus on the second term of \ref{appeq8}:
\begin{align}
\gamma^j_{~b}\na_i A^i_{~j} = \na_i(\gamma^j_{~b}A^i_{~j}) - (\na_i \gamma^j_{~b})A^i_{~j}~.
\end{align}
Insertion of the expression for $A^i_{~j}$, it yields
\begin{align}
\gamma^j_{~b}\na_i A^i_{~j} = \phi K a_b +\phi a^i\na_i s_b - a_b (s^j\na_j\phi)-\frac{1}{2}\na^i\phi\na_i s_b~.
\label{appeq10}
\end{align}
Use of \ref{appeq9} and \ref{appeq10} into \ref{appeq8}, we have
\begin{align}
-\frac{1}{2}\gamma^j_{~b}\na_i t^i_{~j\rm{(Jor)}}= D_i T^{i\rm{(BY)}}_b - \frac{1}{8\pi}\Big[\phi a^i\big(\underbrace{K_{ib}+\na_is_b}_{s_i a_b}\big) - \frac{1}{2}\na^i\phi (\na_i s_b)\Big]
\label{appeq11}
\end{align}
which leads to the final expression \ref{appeq12}.

\subsection{Null hypersurface}

For the null case, $ t^{i \rm (null)}_{~j\rm{(Jor)}}$ is given by \ref{BB13}. Here we will use the form given in the second equality of \ref{BB13}.
Taking the covariant derivative, i.e. $\nabla_i  t^{i \rm (null)}_{~j\rm{(Jor)}}$:
\begin{align}
\nonumber
\nabla_i  t^{i \rm (null)}_{~j\rm{(Jor)}} &= \frac{4}{16\pi} \left[ \nabla_i \phi \left( \nabla_j l^i - \delta^i_j \nabla_a l^a \right) + \phi \nabla_i \left( \nabla_j l^i - \delta^i_j \nabla_a l^a \right) - \nabla_j \left( l^b \nabla_b \phi \right) - \frac{1}{2} \nabla_i \left( l_j \nabla^i \phi \right) \right] \\
&= \frac{1}{4\pi} \left[ \nabla_i \left( \phi \nabla_j l^i \right) - \nabla_j \left( \phi \nabla_i l^i \right) \right] - \frac{1}{4\pi} \nabla_j (l^i \nabla_i \phi) - \frac{1}{8\pi} \nabla_i (l_j \nabla^i \phi)~.
\label{Ref2}
\end{align}
Using \ref{nullcon} and it's trace $\na_i l^i= \theta+ \kappa$, in \ref{Ref2} we obtain
\begin{align}
\nabla_i  t^{i \rm (null)}_{~j\rm{(Jor)}} &= \frac{1}{4\pi} \nabla_i \left[ \phi \left( \theta^i{}_j + l^i \omega_j - \delta^i_j (\theta + \kappa) \right) - \delta^i_j (l^a \nabla_a \phi) \right]- \frac{1}{4\pi} \nabla_i \left( \phi l_j k^m \nabla_m l^i \right) - \frac{1}{8\pi} \nabla_i (l_j \nabla^i \phi)~.
\end{align}
Replacing $\delta^i_j = \Pi^i{}_j - k^i l_j$ into the respective terms yields
\begin{align}
\nabla_i t^{i \rm (null)}_{~j\rm{(Jor)}}&=   \frac{1}{4\pi} \nabla_i \left[ \phi \left( \theta^i{}_j + l^i \omega_j - \Pi^i{}_j (\theta + \kappa) \right) - \Pi^i{}_j (l^a \nabla_a \phi) \right] \nonumber \\
&\quad + \frac{1}{4\pi} \nabla_i \left( \phi k^i l_j (\theta + \kappa) \right) + \frac{1}{4\pi} \nabla_i \left( k^i l_j l^a \nabla_a \phi \right) \nonumber 
 - \frac{1}{4\pi} \nabla_i \left( \phi l_j k^m \nabla_m l^i \right) - \frac{1}{8\pi} \nabla_i (l_j \nabla^i \phi)~.
\label{decom}
\end{align}
We identify the terms within the bracket in the first term with the null Brown--York tensor
$T^{i\rm(null)}_{~j}$, defined in \ref{STBY2}. Also denote $A^i{}_j$ as
\begin{equation}
A^i{}_j = \underbrace{\left( \phi k^i (\theta + \kappa) + k^i (l^a \nabla_a \phi) - \phi k^m \nabla_m l^i - \frac{1}{2} \nabla^i \phi \right)}_{B^i} l_j~.
\label{AIJ}
\end{equation}
Then we have
\begin{align}
\nabla_i  t^{i \rm (null)}_{~j\rm{(Jor)}} &= 2 \nabla_i T^{i\rm (null)}_{~~j}+ \frac{1}{4\pi} \nabla_i A^i{}_j~.
\label{main}
\end{align}
Further, taking the projection of \ref{main} by multiplying with $\Pi^j{}_b$, we have
\begin{align}
\frac{1}{2} \Pi^j{}_b \nabla_it^{i \rm (null)}_{~j\rm{(Jor)}} &= \Pi^j{}_b \nabla_i T^{i\rm (null)}_{~~j} + \frac{1}{8\pi} \Pi^j{}_b \nabla_i A^i{}_j~.
\label{main2}
\end{align}

Let us consider each term in \ref{main2} separately for convenience. Looking at the last term $ \Pi^j{}_b \nabla_i A^i{}_j$, we have
\begin{align}
 \Pi^j{}_b \nabla_i A^i{}_j &= \nabla_i (A^i{}_j \Pi^j{}_b) - (\nabla_i \Pi^j{}_b) A^i{}_j~.
 \label{BBB1}
\end{align}
Using \ref{AIJ}, $\Pi^j_{~b}= \delta^j_b + k^j l_b$, and the normalization $l^j k_j=-1$, \ref{BBB1} simplifies to,
\begin{align}
 \Pi^j{}_b \nabla_i A^i{}_j =  - (\nabla_i k^j) l_b A^i{}_j - (\nabla_i l_b) k^j A^i{}_j
= (\nabla_i l_j) k^j B^i l_b + (\nabla_i l_b) B^i~.
\end{align}
Again, use of \ref{nullcon}, and the fact that $\theta_{ij} k^j = 0$, we have
\begin{align}
 \Pi^j{}_b \nabla_i A^i{}_j &= -\left( \omega_i + l_i (k^m \nabla_m l_j) k^j \right) B^i l_b + (\nabla_i l_b) B^i~.
\end{align}
Plugging in the expression of $B^i$, given by \ref{AIJ}, and using $\omega_i k^i = 0$, the above equation reduces to
\begin{align}
 \Pi^j{}_b \nabla_i A^i{}_j &= \cancel{\phi \omega_i (k^m \nabla_m l^i) l_b} + \frac{1}{2}\cancel{(\omega_i \nabla^i \phi) l_b} + \phi (k^m \nabla_m l_j) k^j (\theta + \kappa) l_b \nonumber \\
&\quad + (k^m \nabla_m l_j) k^j l_b (l^a \nabla_a \phi) + \frac{1}{2} (l^i \nabla_i \phi) (k^m \nabla_m l_j) k^j l_b \nonumber \\
&\quad - \theta_{ib} \phi (k^m \nabla_m l^i) - \frac{1}{2} \theta_{ib} (\nabla^i \phi) -\cancel{\phi \omega_i (k^m \nabla_m l^i)l_b} - \frac{1}{2}\cancel{(\omega_i \nabla^i \phi) l_b} \nonumber \\
&\quad + \phi (\theta + \kappa) (k^m \nabla_m l_b) + (k^m \nabla_m l_b)(l^a \nabla_a \phi) + \frac{1}{2} (l^i \nabla_i \phi)(k^m \nabla_m l_b)~.
\label{BBB2}
\end{align}
Replacement of terms involving $(k^m \nabla_m l_j) k^j l_b = - k^m \nabla_m l_b - k^m \nabla_m (k^j l_b) l_j$ in \ref{BBB2}, we have
\begin{align}
\Pi^j{}_b \nabla_i A^i{}_j&= -\cancel{\phi (k^m \nabla_m l_b) (\theta + \kappa)}- k^m \nabla_m (k^j l_b) \phi (\theta + \kappa) l_j - \cancel{(l^a \nabla_a \phi)(k^m \nabla_m l_b)} - (l^a \nabla_a \phi) k^m \nabla_m (k^j l_b) l_j \nonumber \\
&\quad - \cancel{\frac{1}{2} (l^i \nabla_i \phi ) (k^m \nabla_m l_b)}- \frac{1}{2} (l^i \nabla_i \phi) k^m \nabla_m (k^j l_b) l_j - \theta_{ib} \phi (k^m \nabla_m l^i) - \frac{1}{2} \theta_{ib} (\nabla^i \phi) \nonumber \\
&\quad + \cancel{\phi (\theta + \kappa) (k^m \nabla_m l_b)} + \cancel{(k^m \nabla_m l_b)(l^a \nabla_a \phi)} + \cancel{\frac{1}{2}(l_i \nabla^i \phi)(k^m \nabla_m l_b)}~.
\end{align}
After a fair pairwise cancellations in the above equation, we obtain
\begin{align}
 \Pi^j{}_b \nabla_i A^i{}_j&= - \phi \theta_{ib} (k^m \nabla_m l^i) - \frac{1}{2} \theta_{ib} (\nabla^i \phi) - \phi (\theta + \kappa) k^m \nabla_m (k^j l_b) l_j \nonumber \\
&\quad - (l^a \nabla_a \phi) k^m \nabla_m (k^j l_b) l_j - \frac{1}{2} (l^i\nabla_i \phi) k^m \nabla_m (k^j l_b) l_j~.
\label{BBB4}
\end{align}

Now turn to the the first term $\Pi^j{}_b \nabla_i T^{i\rm (null)}_{~~j}$ of \ref{main2}:
\begin{align}
\Pi^j{}_b \nabla_i T^{i\rm (null)}_{~~j} &= \Pi^j{}_b \delta^m_n \nabla_m T^{n\rm (null)}_{~~j} \nonumber \\
&= \Pi^j{}_b (\Pi^m_n - k^m l_n) \nabla_m T^{n\rm (null)}_{~~j} \nonumber \\
&= \Pi^j{}_b \Pi^m_n \nabla_m T^{n\rm (null)}_{~~j}- \Pi^j{}_b k^m l_n \nabla_m T^{n\rm (null)}_{~~j} \nonumber \\
&= \mathcal{D}_i T^{i\rm (null)}_{~~b} - \Pi^j{}_b (\nabla_m T^{n\rm (null)}_{~~j})(k^m l_n) \nonumber \\
&= \mathcal{D}_i T^{i\rm (null)}_{~~b} - \Pi^j{}_b \left[\cancelto{0}{\nabla_m (T^{n\rm (null)}_{~~j} k^m l_n)} - \cancelto{0}{(\nabla_m k^m) T^{n\rm (null)}_{~~j}l_n} - (\nabla_m l_n) T^{n\rm (null)}_{~~j} k^m \right]
\nonumber
\\
&= \mathcal{D}_i T^{i\rm (null)}_{~~b} + (\nabla_j l_i) T^{i\rm (null)}_{~~b} k^j~.
\label{BBB5}   
\end{align}
Further, on using \ref{nullcon}, \ref{STBY2}, and the fact that, $\theta_{ij} k^j = 0$, $\omega_{j} k^j = 0$, and $l_j k^j = -1$, the last term of the \ref{BBB5} yields,
\begin{align}
 \Pi^j{}_b \nabla_i T^{i\rm (null)}_{~~j}  = \mathcal{D}_i T^{i\rm (null)}_{~~b} + \frac{1}{8\pi} \left[ \phi (k^m \nabla_m l_i) \theta^i{}_b - \phi(\theta + \kappa)\Pi^i{}_b (k^m \nabla_m l_i)  - (k^m \nabla_m l_i) \Pi^i{}_b (l^a \nabla_a \phi) \right]~.
 \label{BBB6}
\end{align}
Finally, collecting all the terms obtained in \ref{BBB4} and \ref{BBB6}
and substituting them into \ref{main2}, we obtain
\begin{align}
\frac{1}{2} \Pi^j{}_b \nabla_i  t^{i \rm (null)}_{~j\rm{(Jor)}}
&= \mathcal{D}_i  T^{i\rm (null)}_{~~b} + \frac{1}{8\pi} \Big[\cancel{\phi (k^m \nabla_m l_i) \theta^i{}_b }-  \phi (\theta + \kappa)\Pi^i{}_b (k^m \nabla_m l_i) - (k^m \nabla_m l_i) \Pi^i{}_b (l^n \nabla_n \phi) \Big] \nonumber \\
&\quad + \frac{1}{8\pi} \left[\cancel{- \theta_{ib} \phi (k^m \nabla_m l^i)} - \frac{1}{2} \theta_{ib} (\nabla^i \phi) -\phi (\theta + \kappa)  k^m \nabla_m (k^j l_b) l_j \right. \nonumber \\
&\qquad\qquad \left. - (l^a \nabla_a \phi) k^m \nabla_m (k^j l_b) l_j - \frac{1}{2} (\nabla_i \phi l^i) k^m \nabla_m (k^j l_b) l_j \right]~.
\label{BBB7}
 \end{align}
Substituting $\Pi^i{}_b (k^m \nabla_m l_i) = - k^m \nabla_m (k^j l_b) l_j$ into \ref{BBB7} yields a further simplification and yields Eq. \ref{Ref1}.



\end{document}